\documentclass[journal]{IEEEtran}

\usepackage{cite}

\ifCLASSINFOpdf
   \usepackage[pdftex]{graphicx}
   \graphicspath{{../pdf/}{../jpeg/}}
   \DeclareGraphicsExtensions{.pdf,.jpeg,.png}
\else
\fi
\UseRawInputEncoding
\usepackage{amsmath}
\usepackage{array}
\usepackage{standalone}
\usepackage{etoolbox}
\newtoggle{arXiv}
\newtoggle{bibrun}
\usepackage{newtxtext, newtxmath}
\usepackage{graphicx}
\usepackage{color, calc}
\usepackage{array, booktabs}
\usepackage{tikz, tikz-3dplot}
\usetikzlibrary{arrows, arrows.meta, calc, positioning, decorations.pathmorphing}
\tikzset{>=Stealth}
\usepackage{siunitx, physics}
\newcommand{\I}{\ensuremath{\mathrm{i}}}

\usepackage{enumitem}
\setlist[description]{labelindent=0pt, leftmargin=\parindent, font=\normalfont\itshape}
\usepackage{url}
\usepackage{subfigure}
\usepackage{textcomp}
\usepackage{eurosym}
\usepackage{xfrac}
\usepackage{pgfplots}
\usepackage{stmaryrd}
\pgfplotsset{compat=1.17}
\usepackage{makecell}
\usepackage[utf8]{inputenc}
\usepackage{hyperref}

\begin{document}

\title{High-resolution ultra-low-field MRI with SNRAware denoising}

\author{\IEEEauthorblockN{
		Teresa~Guallart-Naval\IEEEauthorrefmark{2}$^,$\IEEEauthorrefmark{3}$^,$\IEEEauthorrefmark{1},
		Hui~Xue\IEEEauthorrefmark{4}$^,$\IEEEauthorrefmark{1},
		Jos\'e~M.~Algar\'in\IEEEauthorrefmark{3}$^,$\IEEEauthorrefmark{5},
		Eli~G.~Castanon\IEEEauthorrefmark{3},
		Jesús~Conejero\IEEEauthorrefmark{3},
		Fernando~Galve\IEEEauthorrefmark{3},
		Chloe~Najac\IEEEauthorrefmark{2},
		Mary~A.~Nassejje\IEEEauthorrefmark{3},
		John~Stairs\IEEEauthorrefmark{4},
		Ruben~van~den~Broek\IEEEauthorrefmark{2},
		Lorena~Vega-Cid\IEEEauthorrefmark{3},
		Michael~Hansen\IEEEauthorrefmark{6},
		Andrew~Webb\IEEEauthorrefmark{2}$^,$\IEEEauthorrefmark{7},
		and~Joseba~Alonso\IEEEauthorrefmark{3}}
	
	\IEEEauthorblockA{\IEEEauthorrefmark{1}These authors have contributed equally to this work.}\\
	\IEEEauthorblockA{\IEEEauthorrefmark{2}Department of Radiology, Leiden University Medical Center (LUMC), Leiden, The Netherlands}\\
	\IEEEauthorblockA{\IEEEauthorrefmark{3}MRILab, Institute for Molecular Imaging and Instrumentation (i3M), Consejo Superior de Investigaciones Cient\'ificas (CSIC) \& Universitat Polit\`ecnica de Val\`encia (UPV), Valencia, Spain}\\
	\IEEEauthorblockA{\IEEEauthorrefmark{4}Microsoft Research, Health Futures, Washington, USA}\\
	\IEEEauthorblockA{\IEEEauthorrefmark{5}PhysioMRI Tech S.L., Paterna, Spain}\\
	\IEEEauthorblockA{\IEEEauthorrefmark{6}Nvidia, Washington, USA}\\
	\IEEEauthorblockA{\IEEEauthorrefmark{7}Department of Bioengineering, University of Illinois Urbana-Champaign, Illinois, USA}
	
	\thanks{Corresponding authors: T. Guallart-Naval (t.guallart\_naval@lumc.nl) \& J. Alonso (joseba.alonso@i3m.upv.es).}
}


\maketitle

\begin{abstract}
Ultra-low-field (ULF, $<$\,0.1\,T) magnetic resonance imaging (MRI) systems offer advantages in cost, portability, and accessibility, but their current utility is still limited by low signal-to-noise ratio (SNR). Deep learning (DL)-based denoising is a potential strategy to mitigate this limitation. In this work, we present a systematic evaluation of a 329-million-parameter DL denoising model trained using the \emph{SNRAware} framework and applied to 88, 72, and 47\,mT data. Through a series of controlled phantom and \emph{in-vivo} experiments, we assessed model performance as a function of spatial resolution, coil impedance matching, readout bandwidth, input noise level, $k$-space undersampling, anatomy, image contrast, and scanner platform, and compared against analytical denoising algorithms. The model consistently increased the effective SNR of ULF acquisitions, enabling images with nominal spatial resolutions comparable to those commonly used in clinical 3\,T protocols. Residual analyses indicated that the model predominantly removed stochastic noise while preserving underlying signal structure. We have found no indication of hallucination. At the same time, the results highlight some constraints: denoising performance remains dependent on the starting SNR of the acquisition, and training-domain mismatch influences behavior under certain artifact conditions. These findings suggest that DL-based denoising can significantly expand the practical capabilities of ULF MRI.
\end{abstract}

\IEEEpeerreviewmaketitle

\section{Introduction}

\IEEEPARstart{N}{oise} in an image refers to random variations of brightness and is generally unwanted because it does not correlate with the spatial properties of the imaged object. In Magnetic Resonance Imaging (MRI), it often increases uncertainty in both diagnostic interpretation and quantitative analysis of medical images \cite{AjaBookNoise}. Noisy images result from reconstructing noisy data in the reciprocal spatial-frequency domain, $k$-space, where signals are electronic in origin. Random voltage fluctuations can stem from thermal processes in the sample or the resonant radio-frequency (RF) receive chain \cite{Webb2023b}, or from electromagnetic interference (EMI) coupled into the system \cite{Biber2025}, all three leading to noisy $k$-space data. Importantly, their manifestation in the reconstructed images differs fundamentally from other confounding effects in MRI, such as aliasing, motion, eddy current distortions, off-resonance phenomena, or gradient non-linearities, which do not exhibit stochastic behavior \cite{Bellon86}.

The factors that determine the signal-to-noise (SNR) in MRI can be influenced at three different stages of the imaging process. Before acquisition, hardware design, grounding, coil engineering, shielding, impedance matching, and EMI suppression can reduce electronic noise, and in the best case the system can operate close to the so-called ``thermal limit'' \cite{Guallart-Naval2026a}. During acquisition, imaging sequence timing, data averaging, bandwidth choices, and encoding strategies further influence the effective noise level \cite{BkHaacke}. Finally, after acquisition, noise can be reduced through dedicated denoising techniques applied either in image space or in $k$-space. The present work focuses on this latter stage, where post-processing strategies aim to recover the underlying signal (or reconstruction) from a noisy measurement.

In image space, classical (analytical) denoising methods consist of local and/or non-local filtering and regularization \cite{Manjon2012,Mohan2014}. In $k$-space, analytical denoising is often implemented through filtering or masking strategies, or through reconstruction approaches that produce denoised images from noisy data \cite{Kyung2018,Wang2025,Ciulla2025}. Deep learning (DL) methods have also been proposed for both image-space and reconstruction-integrated denoising, and have shown substantial improvements over traditional approaches in a wide range of MRI applications \cite{Manjon2018,Ongie2020}. The performance of DL-based denoisers depends strongly on their training strategy. For this reason, SNRAware was recently introduced as a model-agnostic training framework, with demonstrated strong results at 1.5\,T and 0.55\,T \cite{Xue2025,Xue2025b}.

Ultra-low-field (ULF) MRI scanners can be made mobile, affordable, and easy to deploy \cite{Webb2023,Guallart-Naval2022,Zhao2024}, although they are also inherently SNR-limited \cite{Sarracanie2020,Webb2023b}. Software denoising is therefore an attractive route to improve image quality \cite{Ayde2025}, and both classical methods (e.g. Block-Matched 4-Dimensional filters, BM4D \cite{Maggioni2013}) and DL approaches \cite{Liu2021,Zhao2024b,Salehi2025,Ilicak2025} have been explored.

In this paper, we conduct a systematic evaluation of a model containing 329 million tunable parameters and trained with SNRAware to determine how it generalizes to the ULF setting. We examine its performance across a range of anatomies, contrasts, spatial resolutions, bandwidth conditions, noise environments, and sampling schemes, and compare it against both standard Fourier reconstructions and classical denoising. Our aim is to determine whether AI trained with high-field data can perform well at ULF-MRI, characterize the present operational limits, identify scenarios in which the model remains reliable, and highlight conditions under which its performance begins to degrade.

\section{Theory framework}
\label{sec:theory}

\subsection{Principles of SNRAware-trained denoising}
\label{sec:denoising_principles}

\begin{figure}
	\centering
	\includegraphics[width=\columnwidth]{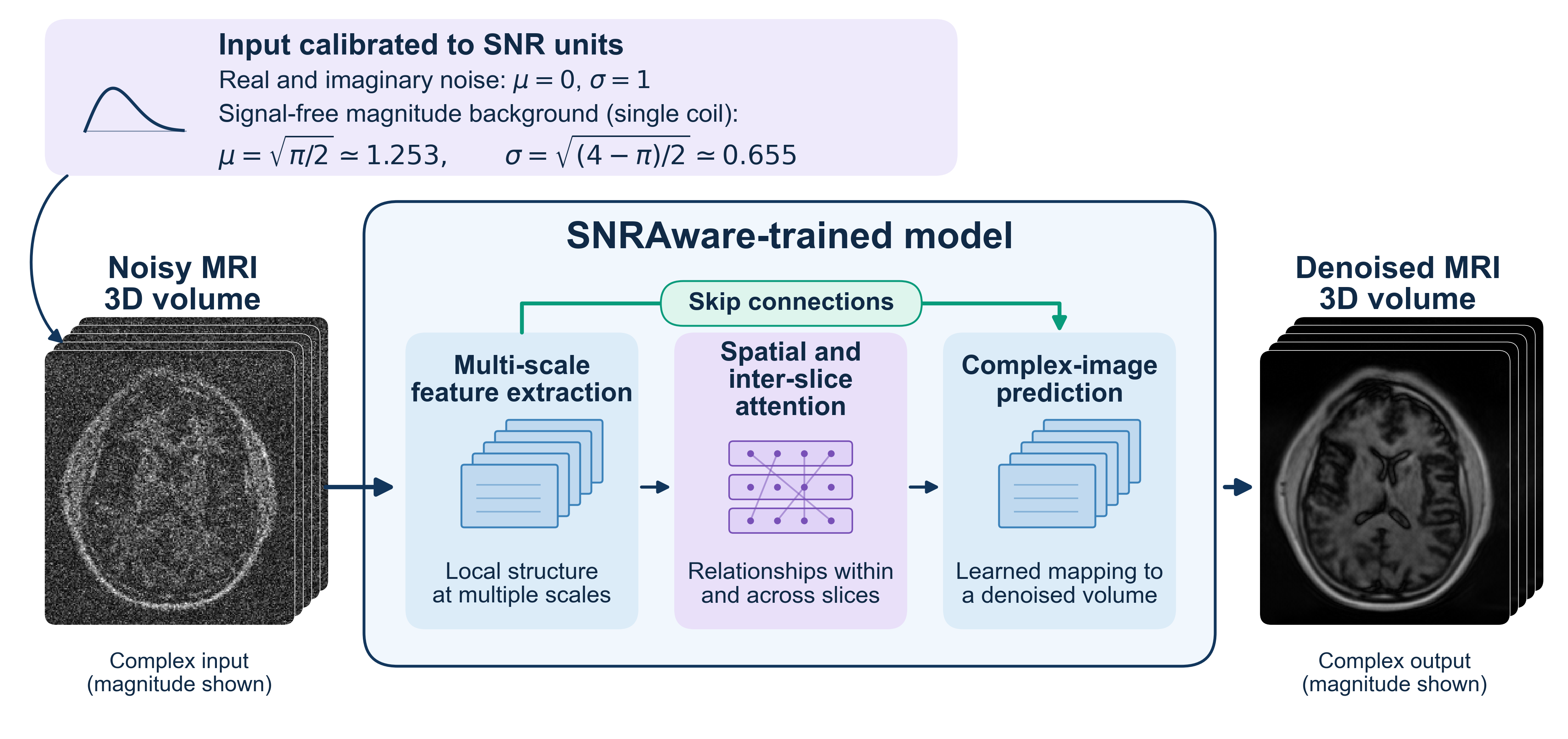}
	\caption{Conceptual illustration of SNRAware-trained denoising (details in Sec.~\ref{sec:denoising_principles}). Crucially, the complex input volume must be calibrated to SNR units (Sec.~\ref{sec:snr_normalization}). The diagram summarizes functional roles rather than the exact sequence of network layers.}
	\label{fig:denoising_concept}
\end{figure}

SNRAware is a model-agnostic training framework that combines learned image representations with quantitative knowledge of MRI noise. Details of the training methodology and the comparison of fourteen network architectures are provided in Ref.~\cite{Xue2025}, while the related imaging-transformer architecture is described in Ref.~\cite{Xue2025b}. Its central principle is that denoising benefits from learning both the statistical properties of noise and the structural relationships present in MR images.

The published SNRAware models were trained using over 2.8 million cardiac cine images acquired at 3\,T \cite{Xue2025}. Cine sequences provide substantial redundancy: anatomical structures move and deform coherently across successive frames, whereas stochastic noise fluctuations are less consistent. Training on these data therefore provides repeated examples of how meaningful image structure relates across frames despite changing noise realizations. Realistic noise simulation, SNR-unit reconstruction, and g-factor augmentation further expose the models to a range of noise levels, spatial variations, and reconstruction-induced correlations.

\begin{figure}
	\centering
	\includegraphics[width=\columnwidth]{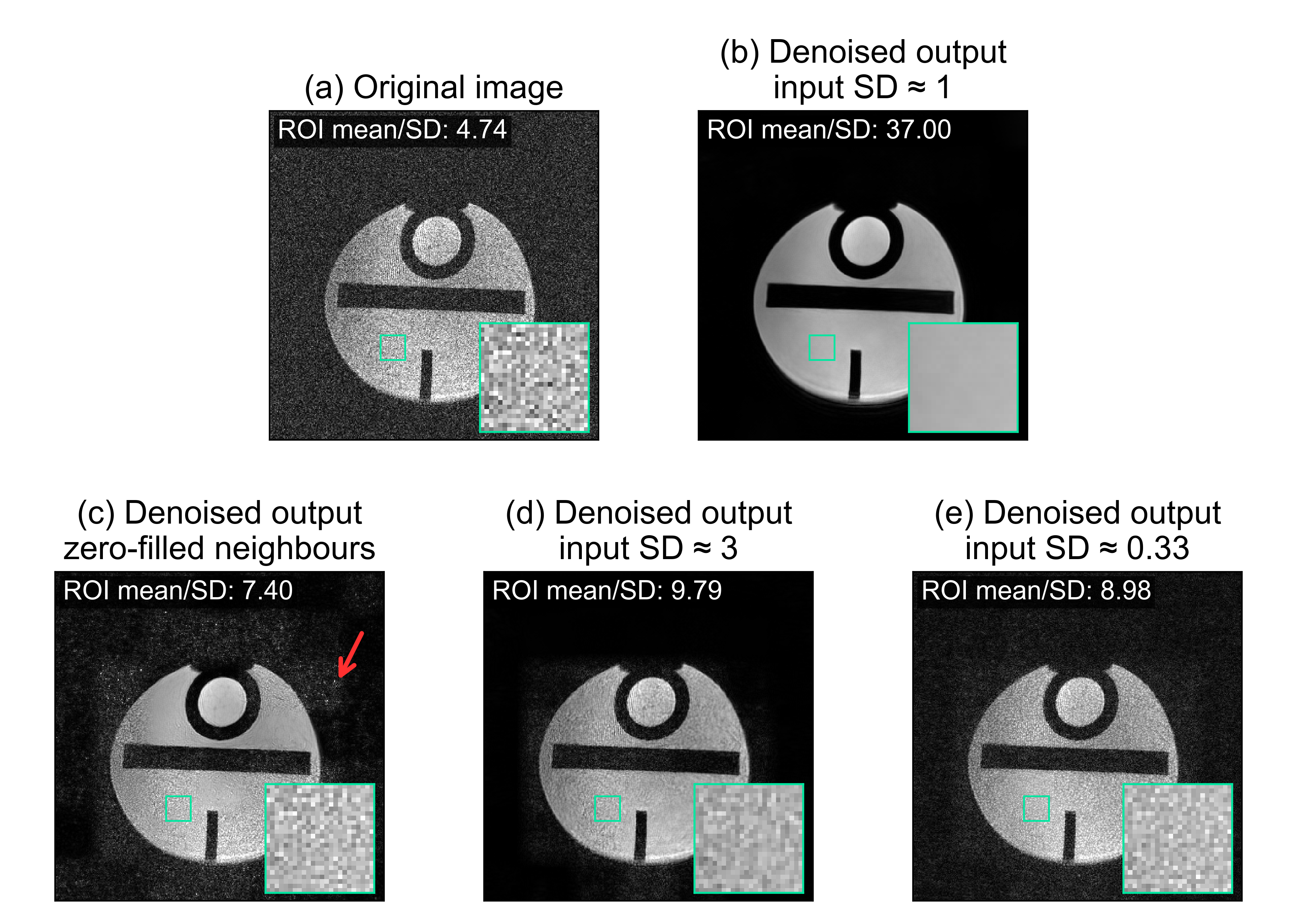}
	\caption{Effects of input normalization and neighbouring-slice information on SNRAware denoising. (a) Original image; (b) full-volume output at unit input noise component SD; (c) output at unit noise SD with all other input slices set to zero; (d,e) full-volume outputs with input noise component SDs of 3 and 0.33, respectively. Green boxes and insets show the common ROI used for the magnitude mean-to-SD ratio; the red arrow marks a noisy background patch.}
	\label{fig:normalization_sensitivity}
\end{figure}

Figure~\ref{fig:denoising_concept} illustrates how these sources of information can contribute to denoising. Convolutional processing extracts features representing local structure and broader spatial context. Attention mechanisms assign importance to relationships between features within images and across frames or slices, while skip connections preserve access to earlier representations. The resulting information is combined to predict a denoised complex image. Noise suppression emerges from this learned mapping, informed jointly by structural context and the noise statistics represented during training.

This combination offers an intuitive explanation for the effectiveness of multi-image denoising. A faint feature that is ambiguous in one image can be interpreted using related information elsewhere in the input. When the input is a three-dimensional volume, neighbouring slices provide an analogous source of redundancy: anatomy changes coherently through the volume, allowing spatial context to support the prediction of structures that are difficult to identify in an individual noisy slice. Temporal coherence in cine data and spatial coherence across slices are not identical, but both provide relationships that a model can exploit beyond the information available at a single location. Consistent with this interpretation, retaining only one input slice and setting all other slices to zero reduces denoising effectiveness compared with processing the complete volume, as illustrated in Fig.~\ref{fig:normalization_sensitivity}b,c.

\subsection{SNR-unit normalization}
\label{sec:snr_normalization}

Appropriate input calibration is crucial to the above process. The significance of an observed intensity fluctuation depends on its amplitude relative to the expected noise level. Incorrect scaling changes this relationship and can therefore alter how the model balances noise suppression against preservation of image structure. SNR-unit normalization establishes the required noise scale.

\begin{figure}
	\centering
	\includegraphics[width=0.8\columnwidth]{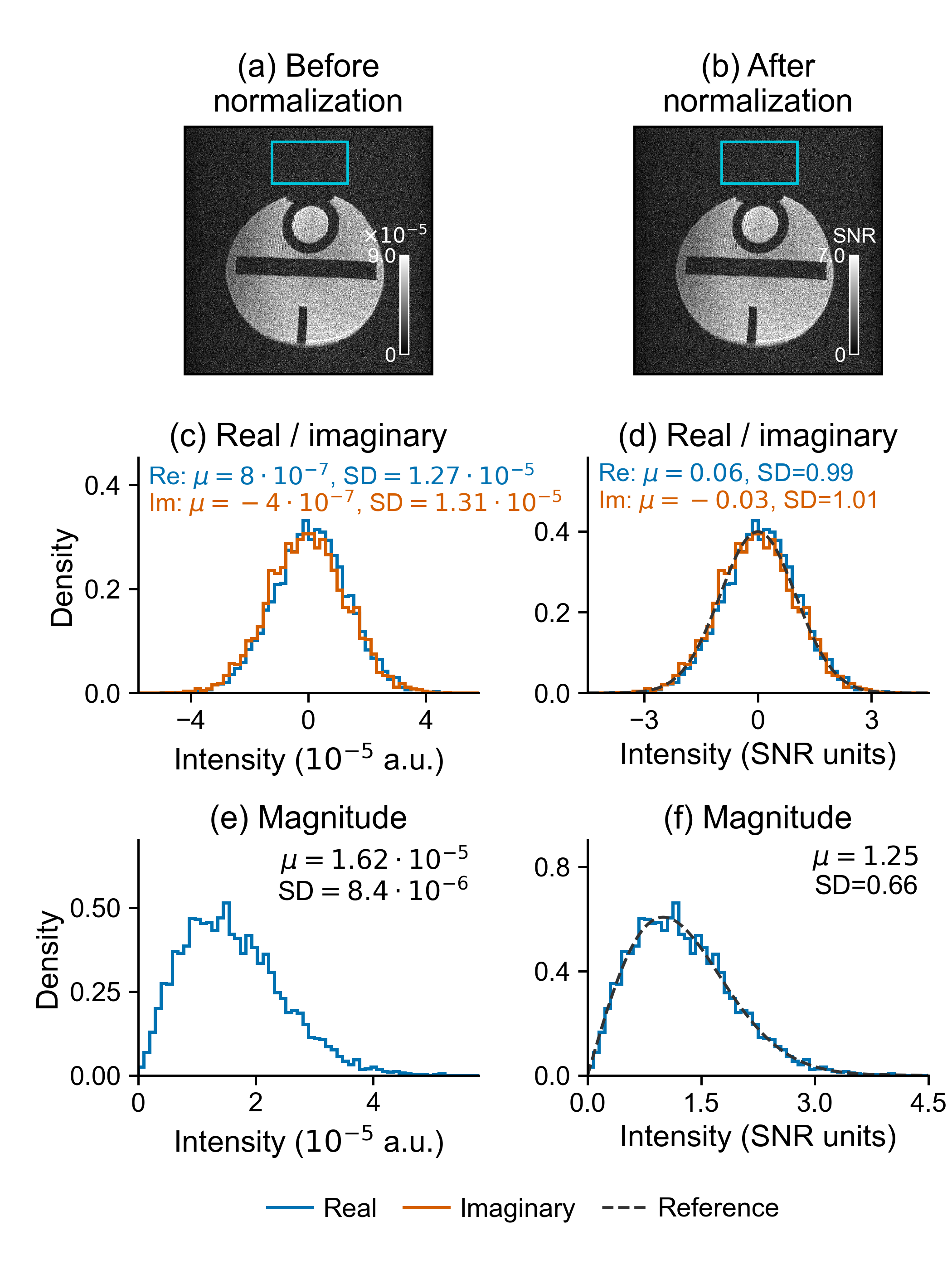}
	\caption{Illustration of SNR-unit normalization using a non-denoised phantom image. The same magnitude slice is shown before (a) and after (b) division of the complex image by the component noise SD, estimated by pooling the real and imaginary variances in the outlined background region. Corresponding background histograms are overlaid for the real and imaginary components (c,d), with magnitude histograms shown separately (e,f). Dashed curves indicate the expected standard Gaussian and unit-scale Rayleigh distributions after normalization.}
	\label{fig:snr_normalization}
\end{figure}

Consider a complex image $\mathcal{I}=\mathcal{S}+\mathcal{N}$, where $\mathcal{S}$ represents the underlying signal and $\mathcal{N}=\mathcal{N}_{\mathrm{re}}+\I\mathcal{N}_{\mathrm{im}}$ is additive noise. At each voxel $\mathbf{r}$, the real and imaginary noise components are assumed to be independent, zero-mean Gaussian variables with equal variance:
\begin{equation*}
	\mathcal{N}_{\mathrm{re}}(\mathbf{r}),
	\mathcal{N}_{\mathrm{im}}(\mathbf{r})
	\sim\mathcal{N}(0,\sigma_{\mathcal{I}}^2).
\end{equation*}
Normalization consists of dividing the complex image by the noise standard deviation (SD) in either component \cite{Kellman2005}:
\begin{equation}
	\mathcal{I}_{\mathrm{SNR}}=\frac{\mathcal{I}}{\sigma_{\mathcal{I}}}.
	\label{eq:norm}
\end{equation}
The resulting real and imaginary noise components each have zero mean and unit SD\footnote{These conditions describe exclusively the noise contribution; they do not imply that signal-containing regions have zero mean or unit SD.}. Figure~\ref{fig:snr_normalization} illustrates these distributions before and after normalization using the same non-denoised image.

In signal-free regions, the normalized magnitude follows a Rayleigh distribution with unit scale parameter. Defining $\mathcal{N}_{\mathrm{SNR}}=\mathcal{N}/\sigma$, its expected mean and SD at each voxel are therefore
\begin{align}
	\nonumber \mathbb{E}\!\left[|\mathcal{N}_{\mathrm{SNR}}(\mathbf{r})|\right]
	&=\sqrt{\frac{\pi}{2}}\approx1.253,\\
	\operatorname{SD}\!\left(|\mathcal{N}_{\mathrm{SNR}}(\mathbf{r})|\right)
	&=\sqrt{\frac{4-\pi}{2}}\approx0.655.
\end{align}

In practice, noise statistics can be estimated from a noise-only measurement acquired with the same receiver settings as the imaging data. When such a measurement is unavailable, acquired $k$-space readout lines sufficiently far from the center may provide an approximation, provided that their signal contribution is negligible relative to the noise.

Regardless of whether noise statistics are estimated from a dedicated noise-only measurement or from peripheral $k$-space readouts, the measured variance must be propagated through the reconstruction to determine the image-domain noise SD ($\sigma_{\mathcal{I}}$). For independent noise samples with real and imaginary component SD $\sigma_k$, an unweighted Fourier reconstruction gives
\begin{equation}
	\sigma_{\mathcal{I}}
	= |c|\,\sigma_k\sqrt{N_{\mathrm{acq}}},
	\label{eq:noise_propagation}
\end{equation}
where $N_{\mathrm{acq}}$ is the number of acquired $k$-space voxels and $c$ is the Fourier-transform normalization factor. For the conventional inverse FFT, $c=1/N_{\mathrm{grid}}$, where $N_{\mathrm{grid}}$ is the total number of elements in the reconstruction grid. Consequently,
\begin{equation}
	\sigma_{\mathcal{I}}
	= \frac{\sigma_k\sqrt{N_{\mathrm{acq}}}}{N_{\mathrm{grid}}}.
	\label{eq:fft_noise_scaling}
\end{equation}
For fully sampled data, this reduces to $\sigma_k/\sqrt{N_{\mathrm{grid}}}$.

Alternatively, when a sufficiently large signal-free image region is available and the noise is spatially uniform, $\sigma_{\mathcal{I}}$ can be estimated directly from the real and imaginary image components.

Figure~\ref{fig:normalization_sensitivity}d,e illustrate the effect of incorrect input normalization. Scaling the same complex volume changes its noise amplitude relative to the distribution expected by the model, while preserving the physical input SNR. Both departures from unit noise SD leave visibly greater fluctuations in the denoised images.

\subsection{Sensitivity profile correction}
\label{sec:sensitivity_correction}

\begin{figure}
	\centering
	\includegraphics[width=0.8\columnwidth]{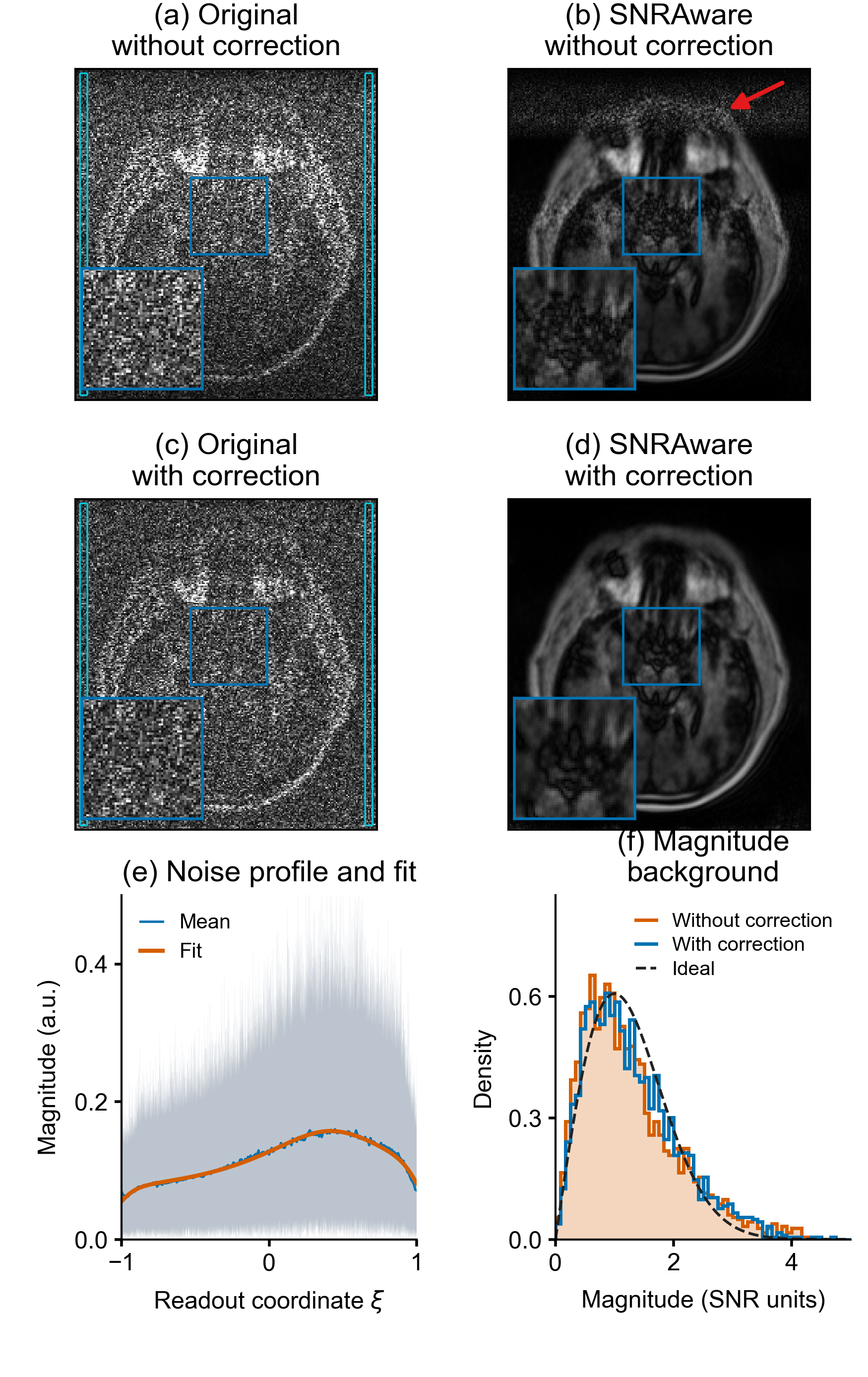}
	\caption{Illustration of receive-chain sensitivity-profile correction using a T$_1$-weighted brain acquisition. Non-denoised originals (a,c) and corresponding SNRAware outputs (b,d) are shown without correction (top) and with correction (middle). The red arrow in (b) indicates a noisy band remaining without correction. (e) Peripheral imaging readouts used as a noise proxy: individual magnitude profiles (light gray), their mean (blue), and a fitted profile following Eq.~(\ref{eq:relative_noise_profile}) (orange). (f) Background magnitude distributions from the non-denoised originals, without correction (orange outline and fill) and with correction (blue outline), compared with the unit-scale Rayleigh distribution (dashed black).}
	\label{fig:sensitivity_profile}
\end{figure}

RF receive coils are resonant circuits and map a frequency-dependent response onto position along the readout direction \cite{Webb2023b}. Amplifiers as well as analog and digital filters in the receive chain further modify this response (Fig.~\ref{fig:sensitivity_profile}e), which can lead to a mismatch between the noise statistics encountered during training and inference, compromising denoising performance (Fig.~\ref{fig:sensitivity_profile}b).

To mitigate this effect, one can estimate the noise amplitude profile and apply its inverse to both the imaging data and the noise measurements used for SNR-unit normalization. Let
\begin{equation}
	\xi=\frac{2(x-x_{\mathrm{c}})}{L_x}\in[-1,1]
\end{equation}
denote position along the readout direction, normalized by the readout field of view $L_x$, with $x_{\mathrm{c}}$ its center. The resonant contribution to the noise power can be approximated by a Lorentzian function\footnote{The LNA can further reshape the measured noise spectrum because its noise figure depends on the source impedance presented by the coil and matching network, which varies across the resonance \cite{Meys1978}. This effect can become appreciable when the readout bandwidth is comparable to or exceeds the resonance width \cite{Zavjalov2019}.}. All in all, the noise power profile can be modeled as
\begin{equation}
	P(\xi)=
	\left[
	B+\frac{A}{1+\left(\frac{\xi-\xi_0}{\Gamma}\right)^2}
	\right]R(\xi),
	\label{eq:noise_power_profile}
\end{equation}
where $A$ is the resonant power amplitude, $\xi_0$ is the resonance center, and $\Gamma$ is the half-width at half-maximum of the Lorentzian contribution. An additive, non-resonant noise floor ($B$) can be included to account for receiver noise that does not follow the coil resonance \cite{Zavjalov2019}. If the underlying edge-filter transfer functions are unknown, an empirical power response can be represented by
\begin{equation}
	R(\xi)=
	\frac{1}{1+\exp[-(\xi+a)/w]}
	\frac{1}{1+\exp[-(a-\xi)/w]},
	\label{eq:edge_response}
\end{equation}
where $a>0$ specifies the symmetric transition locations and $w>0$ controls their width

Since the acquired signals are voltages, the corresponding root-mean-squared (RMS) amplitude follows the square root of the power profile:
\begin{equation}
	q(\xi)=
	\frac{\sqrt{P(\xi)}}
	{\left\langle\sqrt{P(\xi)}\right\rangle_{\xi}},
	\label{eq:relative_noise_profile}
\end{equation}
where the denominator is the arithmetic mean over the sampled readout coordinates. The corrected complex image is then
\begin{equation}
	\mathcal{I}_{\mathrm{corr}}(\mathbf{r})
	=\frac{\mathcal{I}(\mathbf{r})}{q(\xi(x))}.
	\label{eq:profile_correction}
\end{equation}
The same divisor is applied to the noise measurements after transformation into the corresponding readout coordinate, or equivalently to their estimated amplitude profile. SNR-unit normalization (Sec.~\ref{sec:snr_normalization}) is subsequently calculated from these corrected noise statistics, ensuring that the image and its noise estimate undergo the same processing. Because $q(\xi)$ is real and positive, this operation preserves image phase while equalizing the noise amplitude. It also compensates receive-related signal shading to the extent that signal and total noise follow the same frequency dependence.

The visibly different denoising outcomes in Fig.~\ref{fig:sensitivity_profile}b,d, despite the similar background magnitude histograms in panel~(f), illustrate the model's sensitivity to spatial noise calibration.

\subsection{Analysis of complex image residuals}
\label{sec:residual_analysis}

A denoiser should suppress noise while preserving the underlying signal. Consequently, an increase in SNR should be accompanied by evidence that meaningful image information has not been removed. Residual analysis can addresses this complementary question by examining the information discarded during denoising \cite{Brunet2009}. For an input complex image $\mathcal{I}$ and its denoised output $\mathcal{O}$, the complex residual is
\begin{equation}
	\mathcal{R}=\mathcal{I}-\mathcal{O}.
	\label{eq:complex_residual}
\end{equation}
Ideally, the output approaches the underlying signal and the residual contains only noise. Under the assumption of additive noise independent of the signal, this ideal separation also implies no correlation between output and residual. In practice, SNR improvement, output-residual correlation, and residual spatial structure constitute three complementary indicators of denoising performance.

Noise suppression can be quantified through the SNR increase within a selected signal region. For magnitude images, a practical estimator is
\begin{equation}
	\widehat{\mathrm{SNR}}(\mathcal{X})
	=
	\frac{\left\langle|\mathcal{X}|\right\rangle_{\Omega_{\mathrm{S}}}}
	{\operatorname{SD}_{\Omega_{\mathrm{B}}}(|\mathcal{X}|)},
	\qquad
	\mathcal{X}\in\{\mathcal{I},\mathcal{O}\},
	\label{eq:roi_snr}
\end{equation}
where $\Omega_{\mathrm{S}}$ and $\Omega_{\mathrm{B}}$ are signal and signal-free background regions, respectively, and SD denotes the sample standard deviation. The SNR boost is the ratio $\widehat{\mathrm{SNR}}(\mathcal{O})/\widehat{\mathrm{SNR}}(\mathcal{I})$, using identical regions before and after denoising. This is an apparent magnitude SNR: nonlinear denoising can affect background fluctuations differently from those within the object. A large boost therefore indicates background noise suppression but does not, by itself, establish anatomical preservation.

Output-residual dependence can be assessed using the centered Pearson correlation, $\rho(a,b)=\operatorname{Cov}(a,b)/[\operatorname{SD}(a)\operatorname{SD}(b)]$. To retain the signed information in the complex images, correlations are calculated separately for their real and imaginary components. For a given slice $s$, the combined score is
\begin{equation}
	C_s=\frac{1}{2}\left(
	\left|\rho\!\left(
	\operatorname{Re}\mathcal{O}_s,
	\operatorname{Re}\mathcal{R}_s
	\right)\right|
	+
	\left|\rho\!\left(
	\operatorname{Im}\mathcal{O}_s,
	\operatorname{Im}\mathcal{R}_s
	\right)\right|
	\right),
	\label{eq:output_residual_correlation}
\end{equation}
where each component image is treated as a vector containing all pixels. Taking absolute values before averaging prevents positive and negative correlations from cancelling. The equal-weight arithmetic mean of $C_s$ across all slices provides a volume-level score. Values near zero indicate weak linear dependence, whereas larger values indicate that spatial variations in the output are also represented in the residual. Such dependence can arise from removed anatomy or from noise shared between output and residual; conversely, a small correlation does not establish statistical independence or exclude structured signal loss.

Residual structure can also be assessed without comparing it with the output. Moran's I measures the tendency of neighbouring pixels to have similar values \cite{Moran1950}. For each slice, let $m_i=|\mathcal{R}_s(\mathbf{r}_i)|$ denote the complex-residual magnitude at pixel $i$, and let $\bar{m}$ be its mean over the $N$ pixels. Then
\begin{equation}
	I_{\mathrm{M},s}
	=
	\frac{N}{W}
	\frac{
		\sum_{i,j}w_{ij}(m_i-\bar{m})(m_j-\bar{m})
	}{
		\sum_i(m_i-\bar{m})^2
	},
	\qquad
	W=\sum_{i,j}w_{ij}.
	\label{eq:residual_moran}
\end{equation}
Here, $w_{ij}$ can be set to 1 for immediately adjacent horizontal or vertical pixels and zero otherwise, with no connections across image boundaries, while the normalization by $N/W$ and the residual variance makes the index independent of overall intensity scaling and avoids a direct increase with pixel count.

Positive Moran indices indicate spatial clustering, as may arise from anatomical structure, coherent artifacts, or brightness modulations in the residual magnitude. Negative indices indicate a tendency for neighbouring values to alternate, as happens when residuals contain pronounced edge-related oscillations, including around anatomical boundaries. On the other hand, values near zero indicate weak spatial autocorrelation. Independent noise provides a near-zero reference, although reconstruction filtering and denoising can introduce correlations.

The slice-wise Moran profile can be summarized by its arithmetic mean and sample SD across the volume. These describe the typical spatial autocorrelation and its variation between slices; the SD is not an uncertainty estimate for the mean. Together with SNR boost and output-residual correlation, these measures help identify noise suppression accompanied by potentially meaningful information loss, while remaining complementary to visual inspection and reference-based validation.

\begin{figure}
	\centering
	\includegraphics[width=\columnwidth]{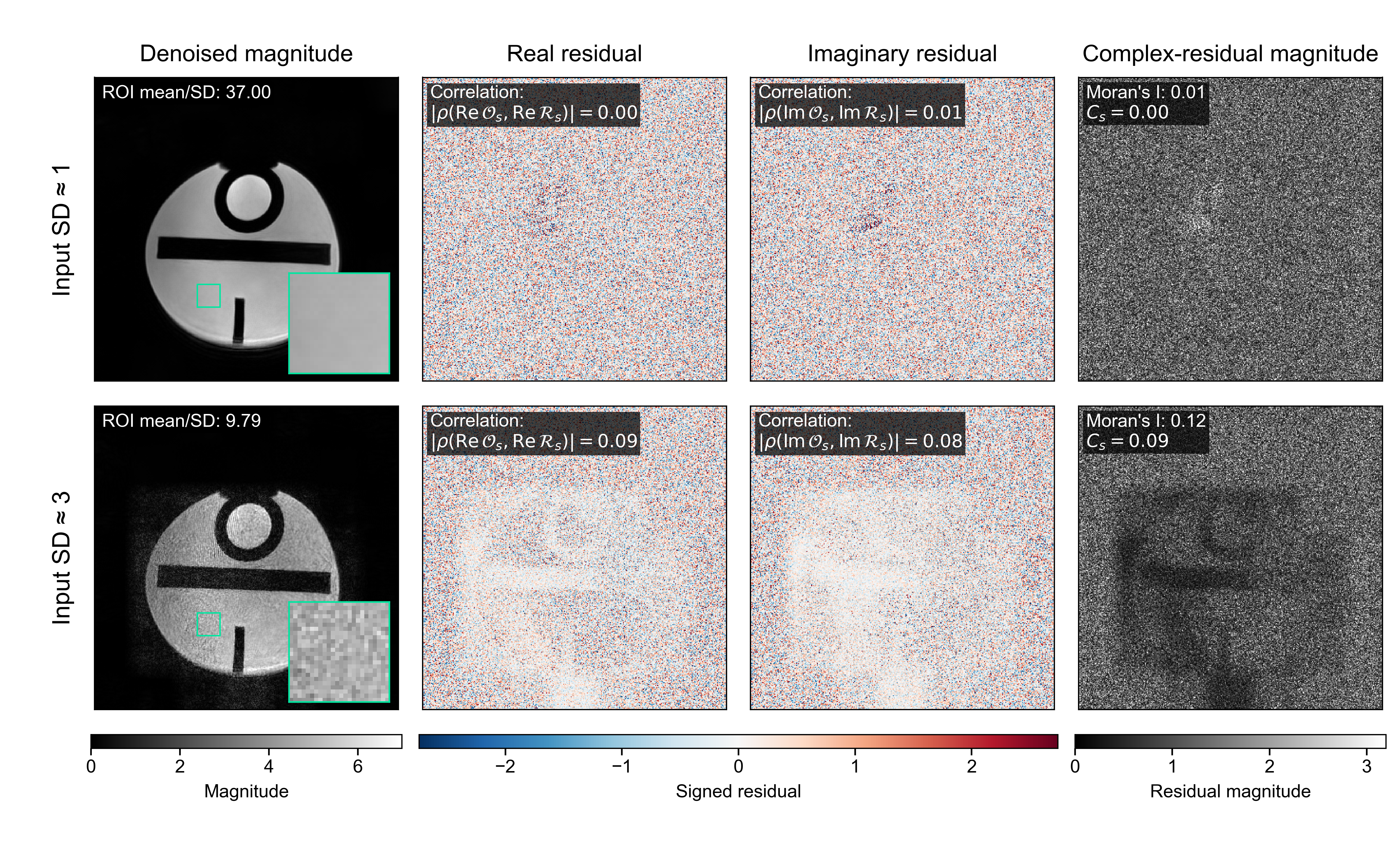}
	\caption{Residual analysis of the same phantom slice with input noise component SDs of 1 (top) and 3 (bottom). Columns show denoised magnitude, real and imaginary residual components, and complex-residual magnitude. Annotations report the magnitude mean-to-SD ratio within the green ROI, absolute component-wise output-residual correlations, combined correlation $C_s$, and Moran's I.}
	\label{fig:residual_assessment}
\end{figure}

Figure~\ref{fig:residual_assessment} illustrates the information provided by these measures. With correct input normalization, the denoised image exhibits smaller fluctuations and the residual has low output-residual correlation and spatial autocorrelation. Increasing the input noise component SD to 3 leaves greater fluctuations in the output and increases both residual metrics, despite using the same noise-profile correction.

\section{Methods}

\subsection{SNRAware model and inference}
\label{sec:model_inference}

We used a pretrained imaging-transformer model supplied by Microsoft Research through its Technology Evaluation Program (TEP). The checkpoint contains approximately 329 million parameters and was trained as described in Sec.~\ref{sec:denoising_principles}. Model weights remained fixed throughout this study; no additional training or fine-tuning was performed using the ULF acquisitions.

The model received reconstructed complex image volumes, with real and imaginary components supplied as separate channels and a g-factor map supplied as an additional channel. All acquisitions used single-channel reception and this map was set to unity. Receive-chain sensitivity-profile correction and SNR-unit normalization were applied before inference, following the principles described in Secs.~\ref{sec:sensitivity_correction} and~\ref{sec:snr_normalization}, respectively.

Inference was performed on a remote GPU using the supplied TorchScript model and reconstruction container. To limit GPU memory requirements, the implementation divides each volume into smaller three-dimensional patches and processes two patches at a time. We retained the supplied patch dimensions of $64\times64$ in-plane pixels and 16 slices. Adjacent patches share 16 pixels along each in-plane direction and 8 slices through the volume; predictions in these overlapping regions are averaged to assemble the denoised complex volume on the original image grid. After SNR-unit normalization, no additional intensity scaling was applied during inference.

\subsection{MRI scanners}
\label{sec:MetSca}

\begin{figure}
  \centering
  \includegraphics[width=1\columnwidth]{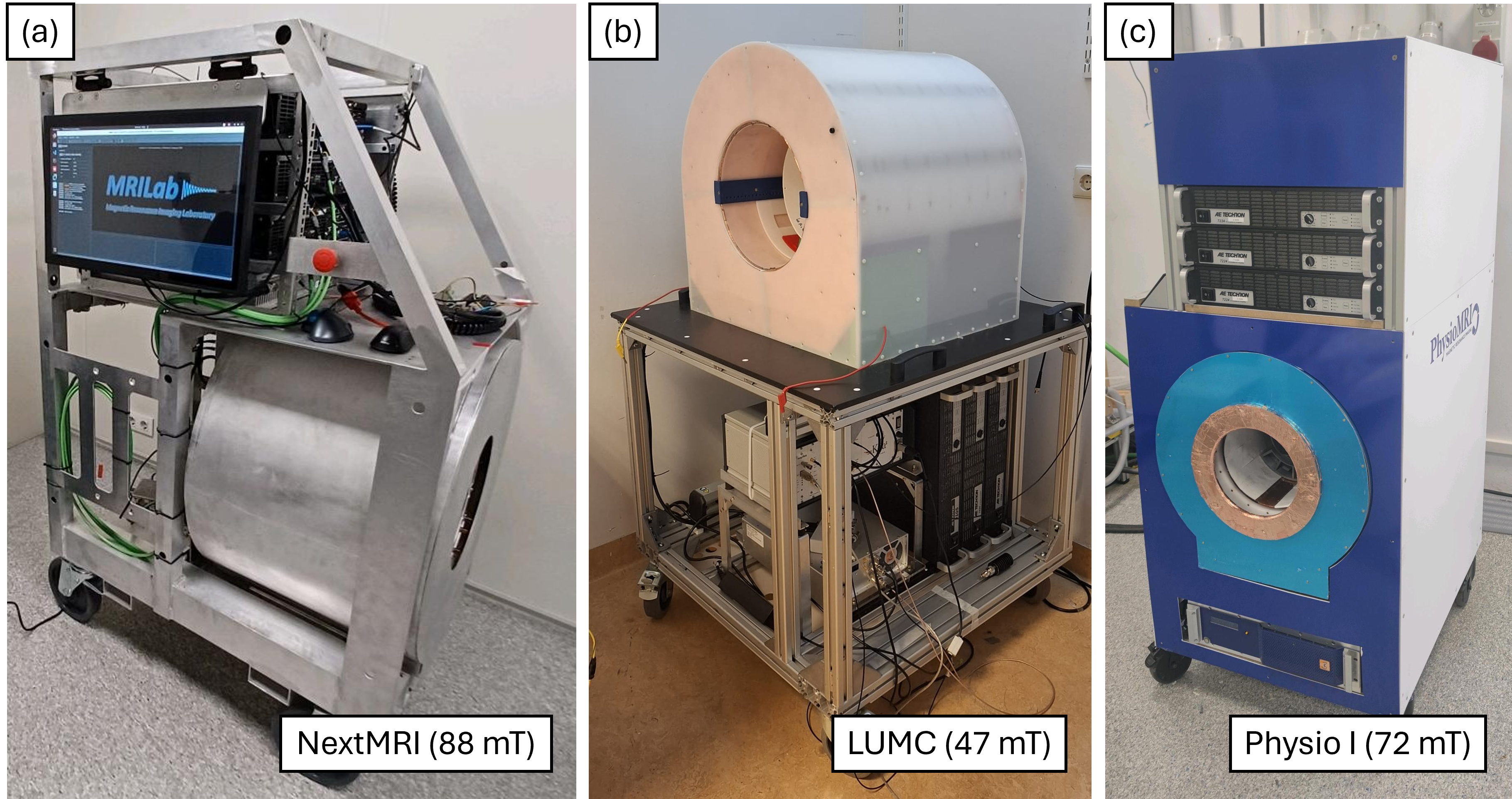}
    \caption{Halbach MRI scanners employed in this work. (a) NextMRI, an 88\,mT elliptical system for head and extremity imaging. (b) The 47\,mT cylindrical brain-imaging system at LUMC. (c) Physio I, a 72\,mT circular system for extremity imaging.}
  \label{fig:scanners}
\end{figure}

Knee, wrist, and most phantom measurements were acquired using NextMRI, a mobile 88\,mT scanner based on an elliptical Halbach-array magnet (Fig.~\ref{fig:scanners}a, \cite{Galve2024}). The homogeneity of the main magnetic field is of $\approx 5,000$\,ppm over a 20\,cm diameter spherical volume (DSV). RF transmission (Tx) and reception (Rx) are performed with combined TxRx coils, and receive signals are routed through a low-noise amplifier chain optimized for operation at the 3.7\,MHz Larmor frequency. System grounding, shielding, and cable routing were configured to minimize EMI during acquisition \cite{Guallart-Naval2026a}, although controlled EMI was introduced in specific experiments to assess model robustness. The RF TxRx coil employed for all knee and phantom images at 88\,mT was a solenoid with diameter and length both 150\,mm. For wrist imaging we employed an elliptical solenoid with inner/outer diameters of 100/140\,mm, and length 200\,mm.

Brain measurements were acquired at Leiden University Medical Center (LUMC) using an in-house 47\,mT cylindrical Halbach scanner (Fig.~\ref{fig:scanners}b, \cite{Lena2026}). The magnet has a field homogeneity of approximately 1,000\,ppm over a 20\,cm DSV. We employed an elliptical spiral-solenoid TxRx head coil measuring 18\,cm in width, 24\,cm in height, and 16\,cm in length, constructed from 15 turns of 1.5\,mm diameter litz wire.

For a comparison between scanners, we acquired additional phantom (American College of Radiology, ACR) measurements using Physio I, a mobile 72\,mT circular Halbach system in Valencia (Fig.~\ref{fig:scanners}c, \cite{Guallart-Naval2022}).

\subsection{System control and data processing}
\label{sec:system_processing}

The NextMRI and Physio I scanners were controlled using MaRCoS \cite{Negnevitsky2023,Guallart-Naval2022b} through the MaRGE graphical interface \cite{Algarin2024}. This workflow integrates sequence execution, data acquisition, and transfer to the remote reconstruction service. The LUMC scanner used a Magritek Kea2 spectrometer.

ULF reconstruction, sensitivity-profile correction, and SNR-unit normalization were performed using SNR-ND, which we have released as an open-source tool \cite{SNRND2026}. SNR-ND also supports inference with the publicly released small, medium, and large SNRAware models. The present study used the separately supplied 329-million-parameter checkpoint, with inference performed through Tyger \cite{Tyger}. The MaRCoS-MaRGE integration enabled processed images to be returned to the scanner console for visualization \cite{Guallart-Naval2026b}. Kea2 acquisitions were exported for subsequent processing through the remote pipeline. Complex non-denoised images and their corresponding denoised outputs were retained for the comparisons and residual analyses.

\subsection{Denoising performance experiments}
\label{sec:MetExps}

\begin{table*}
\caption{Ultra-low-field acquisition parameters for the RARE experiments used in phantom and \emph{in-vivo} studies.}
\centering
\resizebox{\textwidth}{!}{%
\begin{tabular}{c c c c c c c c c c c c c c c}
\toprule
\shortstack{Experiment\\(Figure)} & \thead{FoV (mm$^3$)} & Matrix &  \thead{PF  (\%)} & \thead{Res. (mm$^3$)} & \thead{BW \\ (kHz)} & \thead{Acq. \\ time (ms)} & \thead{TR \\(ms)} & \thead{TE \\(ms)} & \thead{ETL} & \thead{TI \\(ms)} & \thead{$k$-space \\filling} & \thead{Scan time \\(min)} & \thead{Plane} \\
\midrule
Exp 1: knee (\ref{fig:Exp1}a) & $160\times160\times180$ & $500\times500\times60$ & 100 & $0.32\times0.32\times3$ & 42 & 12 & 200 & 20 & 5 & -- & center-out & 20.0 & AX\\
\midrule
Exp 1: brain (\ref{fig:Exp1}b) & $225\times205\times190$ & $225\times205\times190$ & 65 & $1\times1\times1$ & 20 & 11.3 & 900 & 20 & 10 & 250 & in-out & 31.4 & AX\\
\midrule
Exp 2: 2 mm (\ref{fig:Exp2}) & $180\times160\times160$ & $90\times 80\times 80$  & 100 & $2\times2\times2$ & 23 & 4 & 400 & 20 & 8 & -- & center-out  & 5.3 & SAG\\
\midrule
Exp 2: 1.5 mm (\ref{fig:Exp2}) & $180\times160\times160$ & $120\times 112\times 112$  & 100 & $1.5\times1.4\times1.4$ & 30 & 4 & 400 & 20 & 8 & -- & center-out  & 10.5 & SAG\\
\midrule
Exp 2: 1 mm (\ref{fig:Exp2}) & $180\times160\times160$ & $180\times 160\times 160$  & 75 & $1\times1\times1$ & 45 & 4 & 400 & 20 & 8 & -- & center-out  & 16 & SAG\\
\midrule
Exp 2: 0.2 mm (\ref{fig:Exp2}) & $150\times150\times150$ & $750\times 750\times 36$  & 100 & $0.2\times0.2\times4.2$ & 63 & 12 & 200 & 20 & 5 & -- & center-out  & 18 & AX\\
\midrule
Exp 3-4 (\ref{fig:Exps3-6}a) & $150\times150\times150$ & $320\times320\times36$ & 100 & $0.47\times0.47\times4.17$ & 27/160 & 12/2 & 200 & 20 & 5 & -- & center-out & 7.7 & AX\\
\midrule
Exp 5 (\ref{fig:Exps3-6}b) & $150\times150\times150$ & $320\times320\times36$ & 100 & $0.47\times0.47\times4.17$ & 40 & 8 & 200 & 20 & 5 & -- & center-out & 7.7 & AX\\
\midrule
Exp 6 (\ref{fig:Exps3-6}c) & $150\times150\times150$ & $500\times 500\times 50$  & 100/60 & $0.3\times0.3\times3$ & 42 & 12 & 200 & 20 & 5 & -- & center-out  & 16.7/10 & AX\\
\midrule
Exp 7: knee T1 (\ref{fig:Exp7}a) & $160\times160\times160$ & $500\times500\times50$ & 100 & $0.32\times0.32\times3.2$ & 42 & 12 & 200 & 20 & 5 & -- & center-out & 16.7 & AX/SAG/COR\\
\midrule
Exp 7: knee STIR (\ref{fig:Exp7}b) & $160\times160\times160$ & $160\times160\times30$ & 100 & $1\times1\times5.33$ & 40 & 4 & 1000 & 20 & 5 & 90 & center-out & 16.0 & AX/SAG/COR\\
\midrule
Exp 7: wrist T1 (\ref{fig:Exp7}c) & $160\times120\times80$ & $400\times300\times26$ & 100 & $0.4\times0.4\times3.08$ & 33 & 12 & 200 & 20 & 5 & -- & center-out & 5.2 & COR\\
\midrule
Exp 7: wrist STIR (\ref{fig:Exp7}c) & $160\times120\times80$ & $200\times150\times26$ & 80 & $0.8\times0.8\times3.08$ & 17 & 12 & 1000 & 20 & 5 & 100/80 & center-out & 10.0 & COR\\
\midrule
\shortstack{Exp 7: brain T$_1$\\reused from Exp 1 (\ref{fig:Exp7}d)} & $225\times205\times190$ & $225\times205\times190$ & 65 & $1\times1\times1$ & 20 & 11.3 & 900 & 20 & 10 & 250 & in-out & 31.4 & AX\\
\midrule
Exp 7: brain T2 (\ref{fig:Exp7}e) & $205\times225\times190$ & $205\times225\times60$ & 60 & $1\times1\times3.17$ & 20 & 10.3 & 2500 & 20 & 15 & -- & linear & 18.1 & AX\\
\midrule
\shortstack{Exp 8: phantom\\88/72\,mT (\ref{fig:Exp8})} & $150\times150\times150$ & $320\times320\times36$ & 100 & $0.47\times0.47\times4.17$ & 27 & 12 & 200 & 20 & 5 & -- & center-out & 7.7 & AX\\
\midrule
Exp 9 (\ref{fig:Exp9}) & \multicolumn{13}{c}{Same low-field T$_1$-weighted brain acquisition as Exp 1 and Exp 7.}\\
\bottomrule

\end{tabular}}
\label{tab:seq_params}
\end{table*}

The main goal of this paper is to assess the impact of various operational aspects on the denoiser model performance. To this end, we acquired images in nine different experiments, which we describe below. 

For the quantitative comparisons, $\mathcal{I}$ denotes the non-denoised complex reconstruction, including PF completion where applicable, after sensitivity-profile correction and SNR-unit normalization, following Eqs.~\eqref{eq:profile_correction} and \eqref{eq:norm}. Noise statistics were estimated from dedicated noise-only readouts where available. For the Kea2 brain acquisitions, peripheral acquired $k$-space readouts served as a proxy for estimating the noise level.

For partial Fourier (PF) acquisitions, missing $k$-space coefficients were estimated locally using a phase-constrained projection onto convex sets (POCS) reconstruction \cite{Haacke1991}, before denoising. Noise scaling produced the complex input volume supplied to the denoiser, without further PF processing in the inference container.

We calculated the complex residual $\mathcal{R}$ using Eq.~\eqref{eq:complex_residual} and displayed its magnitude. Apparent magnitude SNR was measured on a selected slice using the same signal and background ROIs for the original and all outputs, as defined in Eq.~\eqref{eq:roi_snr}. The output-residual correlation $C_s$ in Eq.~\eqref{eq:output_residual_correlation} was calculated over all pixels in each slice and averaged with equal weight across slices. Moran's I was calculated on $|\mathcal{R}|$ using Eq.~\eqref{eq:residual_moran} with horizontal and vertical neighbours. We report its mean and sample SD across slices. Where $k$-space is shown, we display the magnitude of the central plane of the 3D Fourier transform of the complex reconstruction on a logarithmic scale.

In Experiments~1 and~9, we compared SNRAware with total variation (TV) regularization and BM4D. TV balances fidelity to the input against a penalty on spatial intensity variations. We used the Berkeley Advanced Reconstruction Toolbox (BART) \cite{Uecker2015} with an in-plane TV penalty and regularization weights $\lambda=0.2$ and $0.5$. BART received the Fourier transform of $\mathcal{I}$, so all three methods operated on the same processed complex reconstruction. BM4D \cite{Maggioni2013} was applied separately to the real and imaginary volumes, using a background noise SD estimated for each component. The filtered components were recombined into a complex output $\mathcal{O}$.

All ULF acquisitions used three-dimensional rapid acquisition with relaxation enhancement (3D RARE) pulse sequences. Sequence parameters and acquisition settings for all subsequent experiments are summarized in Table~\ref{tab:seq_params}.

\subsubsection{3\,T clinical-resolution benchmark}
\label{sec:MetExp1}

Experiment~1 evaluated denoising of ULF acquisitions at nominal spatial resolutions comparable to those used in clinical 3\,T protocols. We used an axial T$_1$-weighted knee acquisition at 88\,mT with $0.32\times0.32\times3$\,mm$^3$ voxels and an inversion-recovery T$_1$-weighted brain acquisition at 47\,mT with 1\,mm isotropic voxels. The same brain acquisition was also used in Experiments~7 and~9.

\subsubsection{Model performance versus spatial resolution}
\label{sec:MetExp2}

Experiment~2 evaluated SNRAware across three spatial resolutions in sagittal T$_1$-weighted knee acquisitions. The voxel sizes were 2\,mm isotropic, $1.5\times1.4\times1.4$\,mm$^3$, and 1\,mm isotropic. We also acquired the ACR phantom at $0.2\times0.2$\,mm$^2$ in-plane resolution, which required gradient amplitudes exceeding 40\,mT/m.

\subsubsection{Model performance versus readout bandwidth}
\label{sec:MetExp3}

Experiment~3 evaluated how readout bandwidth affects denoising performance. For a fixed number of readout samples, a shorter readout window covers a wider frequency range and can expose more of the receive chain's frequency-dependent sensitivity. We acquired the ACR phantom with readout windows of 2\,ms (160\,kHz) and 12\,ms (27\,kHz). Each readout setting was acquired under both coil matching conditions described in Sec.~\ref{sec:MetExp4}, allowing bandwidth to be compared at fixed matching. The remaining sequence parameters were held fixed.

\subsubsection{Model performance versus impedance matching}
\label{sec:MetExp4}

Experiment~4 evaluated whether denoising performance was sensitive to coil impedance matching. Changing the match can alter both the received signal level and the shape of the receive sensitivity profile across the acquisition bandwidth. We compared an optimized match, with reflected power below $-20$\,dB relative to incident power, with a deliberately degraded match of approximately $-3$\,dB. Using the acquisitions from Experiment~3, we compared the two matching conditions separately at each readout bandwidth.

\subsubsection{Model performance versus input noise level}
\label{sec:MetExp5}

Experiment~5 evaluated the robustness of the denoiser to increased environmental noise under otherwise identical imaging conditions. We first acquired the ACR phantom with the system optimized to operate near its thermal noise level. We then placed a switched-mode power supply near the receive coil and repeated the acquisition, producing an approximately fivefold increase in the measured RMS noise voltage. The sequence parameters were held fixed.

\subsubsection{Model performance versus undersampling}
\label{sec:MetExp6}

Experiment~6 evaluated the response of the denoising model to reduced $k$-space sampling. Undersampling can shorten acquisition times but the missing $k$-space data and their treatment during reconstruction can alter the noise structure and introduce artifacts. Two sampling schemes were compared with the ACR phantom: a fully sampled acquisition and a Partial Fourier acquisition with 60\,\% filling in the second phase-encoding direction. All other sequence parameters were kept constant. The PF acquisition was reconstructed using the local POCS procedure described above before applying SNRAware.

\subsubsection{Model performance versus anatomy, orientation, contrast, and scanner}
\label{sec:MetExp7}

Experiment~7 evaluated the generalization of the denoising model across anatomical targets, imaging orientations, contrasts, and scanners. To this end, we acquired T$_1$-weighted and STIR (Short-Tau Inversion Recovery, often employed for fat signal suppression at low field strengths \cite{Ghazinoor2007}) knee images in axial, sagittal, and coronal orientations at 88\,mT. We also acquired coronal wrist images with T$_1$ weighting and STIR preparation, using inversion times of 100\,ms and 80\,ms for the latter. To extend the comparison to brain anatomy, we included inversion-recovery T$_1$-weighted images at 1\,mm isotropic resolution and T$_2$-weighted images at $1\times1\times3.17$\,mm$^3$, acquired at 47\,mT. For each brain acquisition, we selected a superior axial slice containing cortical structures and sulcal CSF spaces and a second slice through the lateral ventricles.

\subsubsection{Model performance versus field strength}
\label{sec:MetExp8}

Experiment~8 compared SNRAware denoising of T$_1$-weighted ACR phantom acquisitions obtained with the NextMRI system at 88\,mT and the Physio~I system at 72\,mT, both described in Sec.~\ref{sec:MetSca}. The acquisitions used the same nominal imaging matrix, field of view, and sequence timing, as summarized in Table~\ref{tab:seq_params}. Differences between the systems, including their RF coils and field homogeneity, may also contribute to the observed differences.

\subsubsection{Comparison with 3\,T clinical references}
\label{sec:MetExp9}

Experiment~9 assessed the correspondence between ULF brain images and clinical-resolution references acquired from the same participant at 3\,T. We used the 47\,mT inversion-recovery T$_1$-weighted acquisition at 1\,mm isotropic resolution, comparing its original reconstruction and the four denoised outputs with T$_1$-weighted, proton-density-weighted, and T$_2$-weighted reference images. A representative axial ULF slice through the lateral ventricles was selected, together with approximately corresponding slices from each 3\,T series.

Each reference was independently registered to the original ULF magnitude image using a two-dimensional rigid transformation comprising in-plane rotation and translation, with the sampling ratio fixed by the recorded pixel spacing. Registration maximized mutual information (MI), a measure of statistical dependence between corresponding image intensities commonly used for multimodality registration \cite{Maes1997}. Reference images were bilinearly resampled onto the ULF grid for metric calculation; the ULF images were not interpolated. For display, the references retained their native pixel resolution and were cropped to the same physical field of view.

MI was calculated from joint intensity histograms using 64 equally spaced bins per image and expressed in bits by using base-2 logarithms \cite{Shannon1948}. All 46,125 pixels of the selected low-field slice were included, without masking or smoothing. The same ULF intensity range and bin edges were used for all methods, with a separate fixed intensity range for each reference contrast. Calculations with 32 and 128 bins were also performed to assess sensitivity to histogram binning.

\section{Results}

\begin{figure*}
	\centering
	\includegraphics[width=1\textwidth]{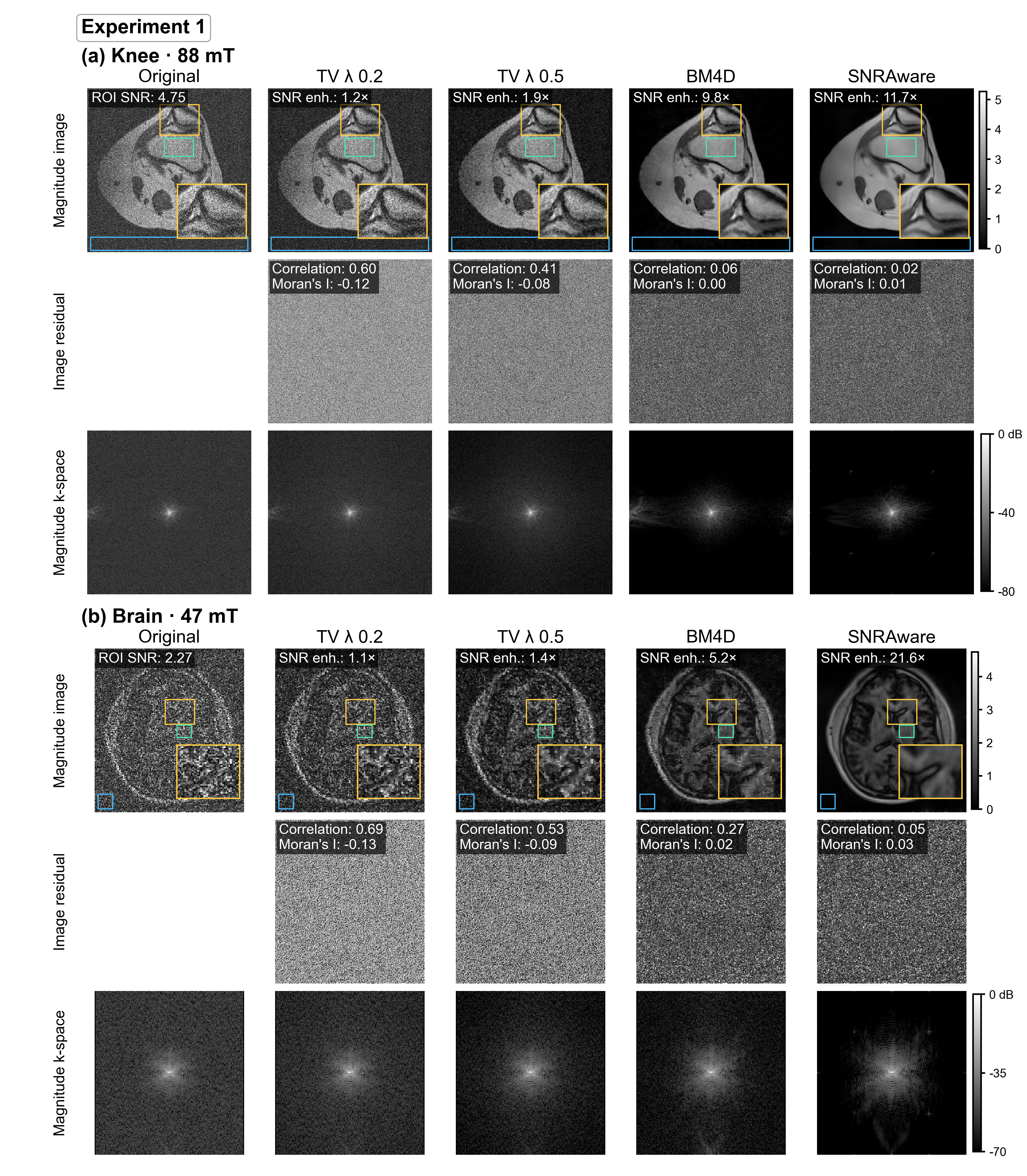}
	\caption{Experiment~1. Representative denoising results for (a) an 88\,mT knee acquisition and (b) a 47\,mT brain acquisition. Columns show the original reconstruction and the outputs of TV regularization ($\lambda=0.2$ and $0.5$), BM4D, and SNRAware. For each acquisition, the rows show magnitude images, complex residual magnitudes $|\mathcal{I}-\mathcal{O}|$, and the magnitude of the central $k$-space plane obtained by a 3D Fourier transform of each complex volume. Yellow boxes mark the magnified regions. Green and blue boxes mark the signal and background ROIs used for the apparent SNR.}
	\label{fig:Exp1}
\end{figure*}

Figure~\ref{fig:Exp1} shows axial knee images acquired at 88\,mT and inversion-recovery T$_1$-weighted brain images acquired at 47\,mT, as described in Sec.~\ref{sec:MetExp1}. For each acquisition, the top row presents the original magnitude image and the corresponding outputs of TV regularization ($\lambda=0.2$ and $0.5$), BM4D, and SNRAware. The middle row displays the complex residual magnitudes $|\mathcal{I}-\mathcal{O}|$, and the bottom row shows the magnitude of the central $k$-space plane of each reconstruction on a logarithmic scale. The green and blue boxes indicate the signal and background ROIs used to calculate the apparent SNR, and the yellow boxes indicate the regions shown in the insets.

\begin{figure}
	\centering
	\includegraphics[width=\columnwidth]{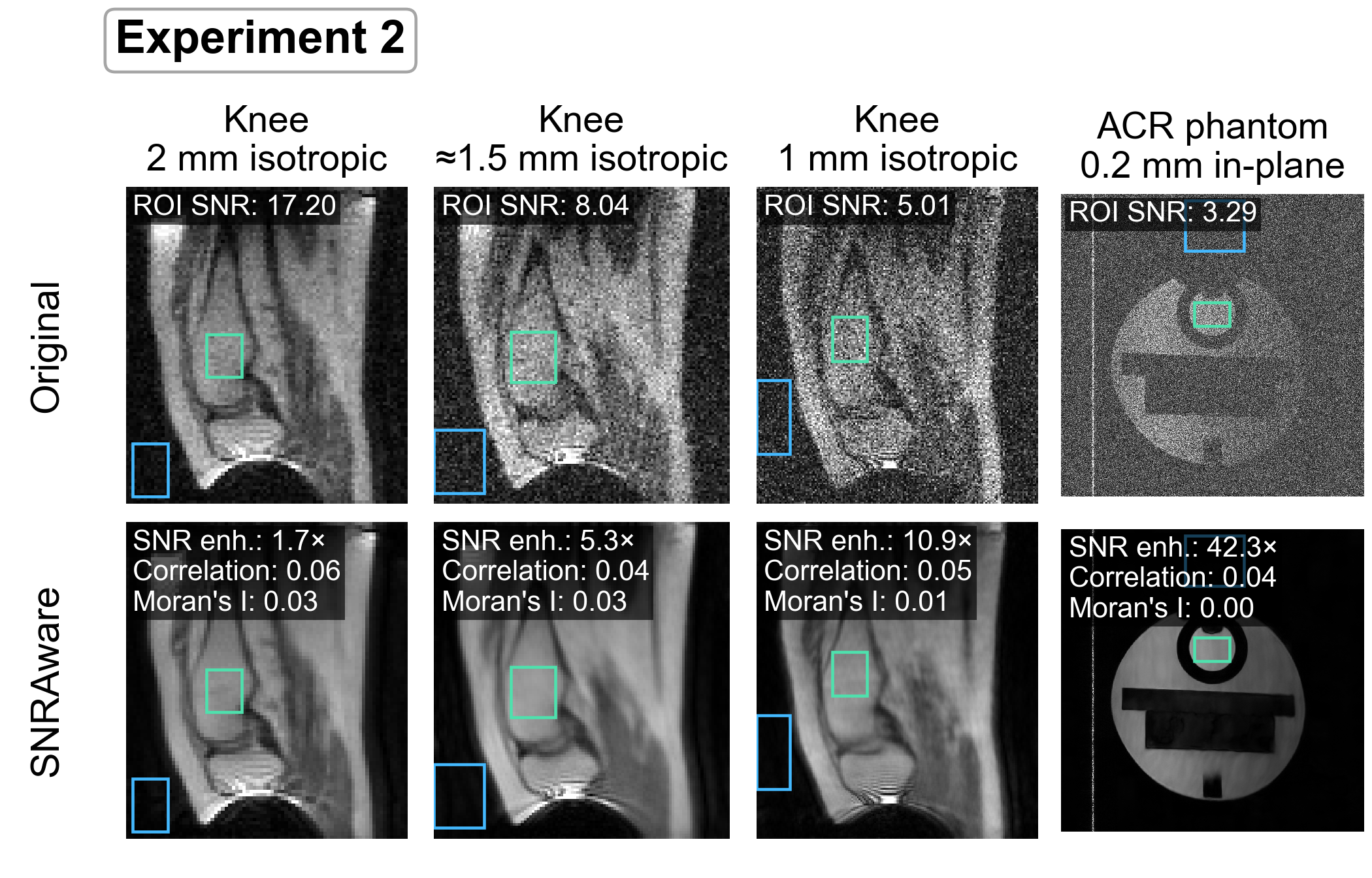}
	\caption{Experiment~2. Sagittal knee acquisitions at nominal voxel sizes of 2, 1.5, and 1\,mm, followed by the ACR phantom acquisition at $0.2\times0.2$\,mm$^2$ in-plane resolution.}
	\label{fig:Exp2}
\end{figure}

Figure~\ref{fig:Exp2} shows representative slices from sagittal knee acquisitions at nominal resolutions of 2\,mm, 1.5\,mm, and 1\,mm, and from an ACR phantom acquisition at an in-plane resolution of $0.2\times0.2$\,mm$^2$, as described in Sec.~\ref{sec:MetExp2}. The top row displays the original magnitude images, and the bottom row shows the corresponding SNRAware-denoised images.

\begin{figure*}
  \centering
  \includegraphics[width=1\textwidth]{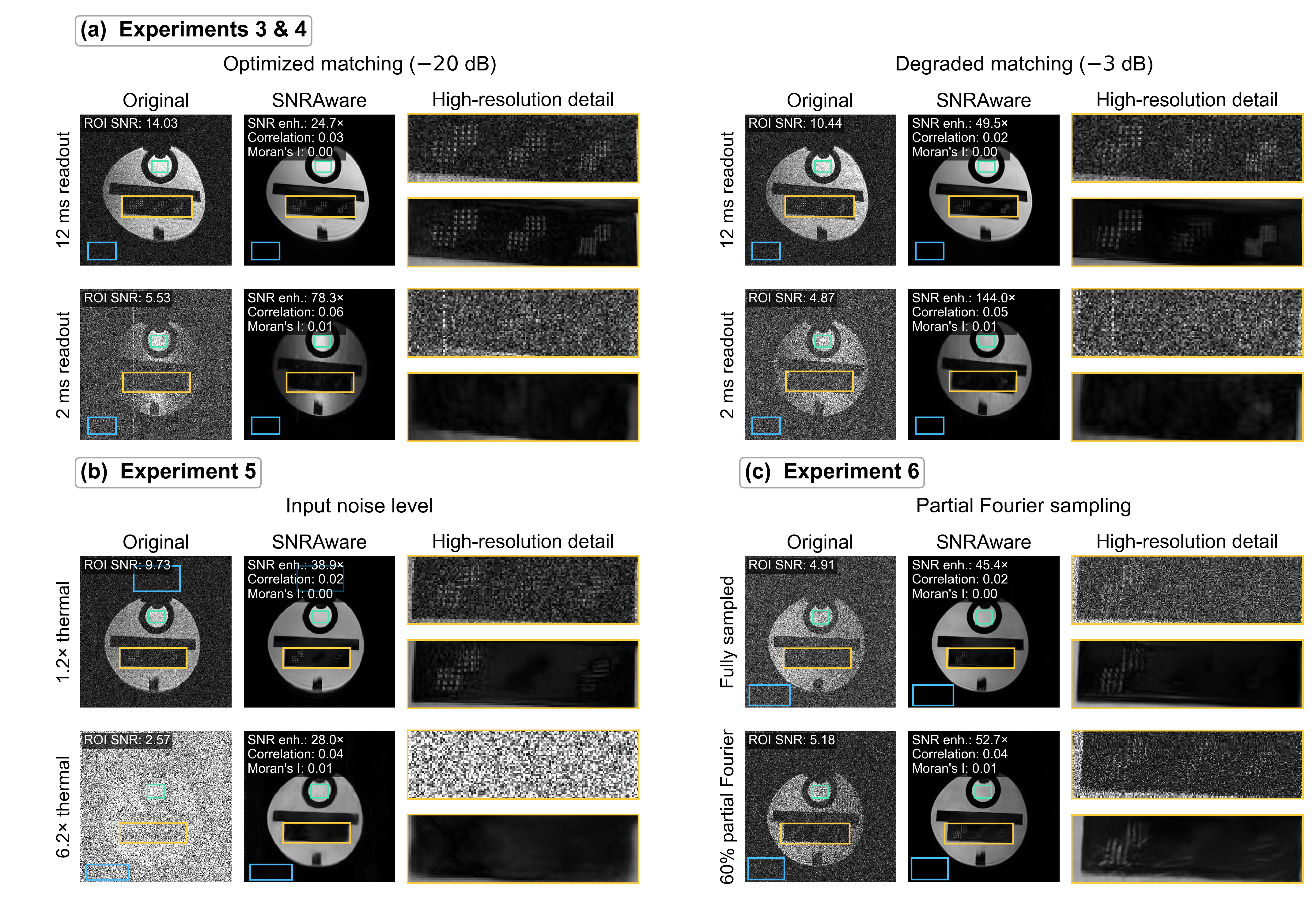}
    \caption{Experiments~3–6. ACR phantom reconstructions comparing original and SNRAware under varying acquisition conditions. (a) Good (–20\,dB) and poor (–3\,dB) impedance matching with 2\,ms and 12\,ms readouts. (b) Baseline (1.2$\times$) and elevated (6.2$\times$) noise levels. (c) Fully sampled and 60\,\% Partial Fourier acquisitions. Zoomed-in panels display the high-resolution region of the ACR phantom.}
  \label{fig:Exps3-6}
\end{figure*}

Figure~\ref{fig:Exps3-6} shows ACR phantom images acquired under the conditions described in Secs.~\ref{sec:MetExp3} through \ref{sec:MetExp6}. Panel~(a) shows 12\,ms and 2\,ms readouts under optimized ($-20$\,dB) and degraded ($-3$\,dB) coil matching. Panel~(b) shows the baseline and elevated-noise acquisitions, and panel~(c) shows the fully sampled and 60\,\% PF acquisitions. Each condition is illustrated by the original magnitude image, the corresponding SNRAware output, and zooms of the high-resolution region.

\begin{figure*}
  \centering
  \includegraphics[width=1\textwidth]{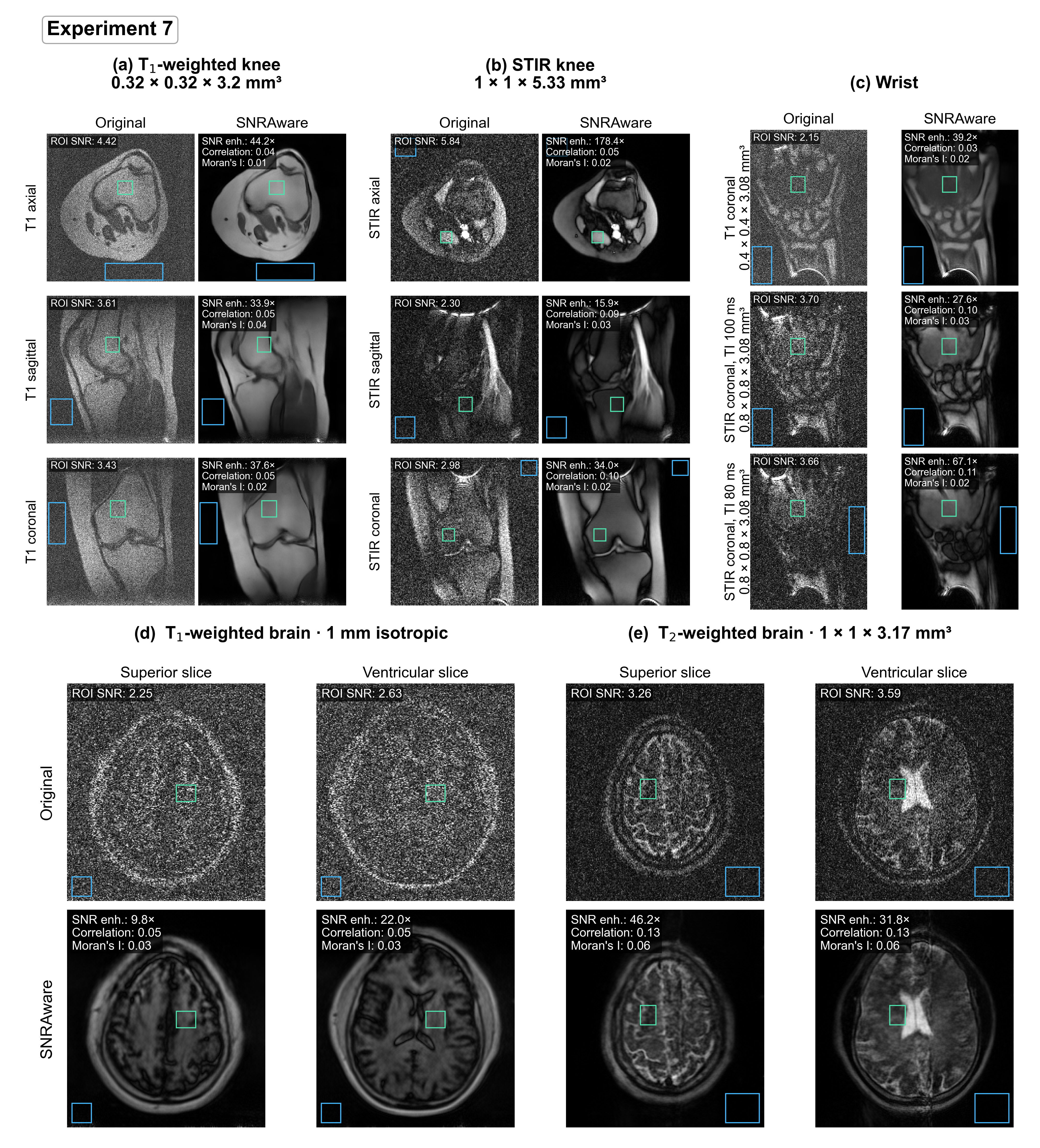}
    \caption{Representative results from Experiment~7. Original magnitude reconstructions and corresponding SNRAware-denoised images acquired under different anatomical and contrast conditions.}
   \label{fig:Exp7}
\end{figure*}

Figure~\ref{fig:Exp7} presents original magnitude images and the corresponding SNRAware outputs under the conditions described in Sec.~\ref{sec:MetExp7} (Experiment~7). Panels~(a) and~(b) show T$_1$-weighted and STIR knee images, respectively, in axial, sagittal, and coronal orientations. Panel~(c) shows coronal wrist images with T$_1$ weighting (top) and STIR preparation with inversion times of 100\,ms (middle) and 80\,ms (bottom). These acquisitions were performed at 88\,mT. Panels~(d) and~(e) show T$_1$-weighted and T$_2$-weighted brain images, respectively, acquired at 47\,mT, with a superior and a ventricular slice for each contrast.

\begin{figure}
	\centering
	\includegraphics[width=\columnwidth]{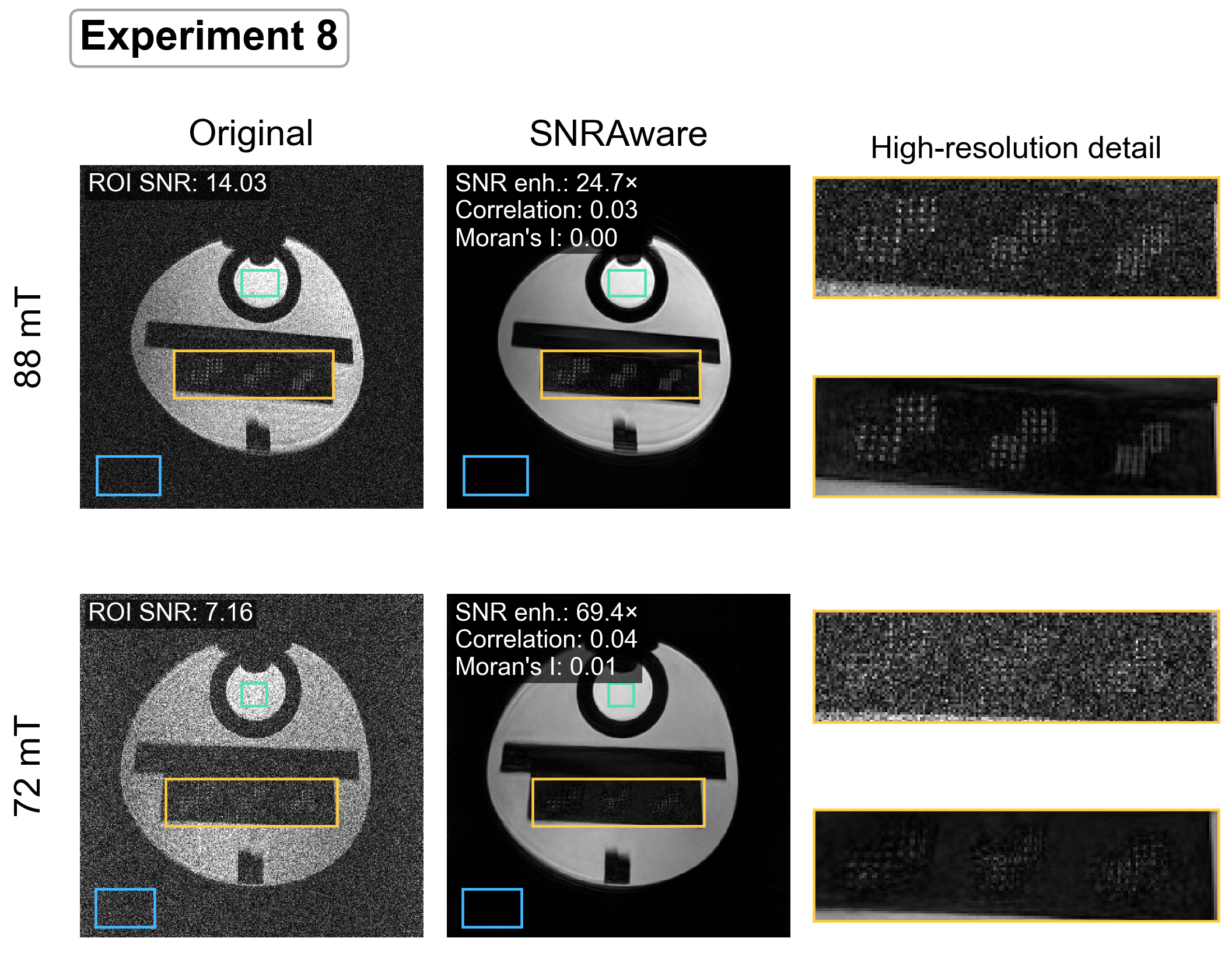}
	\caption{Experiment~8. Original and SNRAware-denoised ACR phantom images acquired at 88\,mT (top) and 72\,mT (bottom), with zooms of the high-resolution region.}
	\label{fig:Exp8}
\end{figure}

Figure~\ref{fig:Exp8} shows representative ACR phantom slices acquired at 88\,mT (top row) and 72\,mT (bottom row). Columns display the original magnitude images, the corresponding SNRAware outputs, and enlarged views of the high-resolution region.

\begin{figure}
	\centering
	\includegraphics[width=\columnwidth]{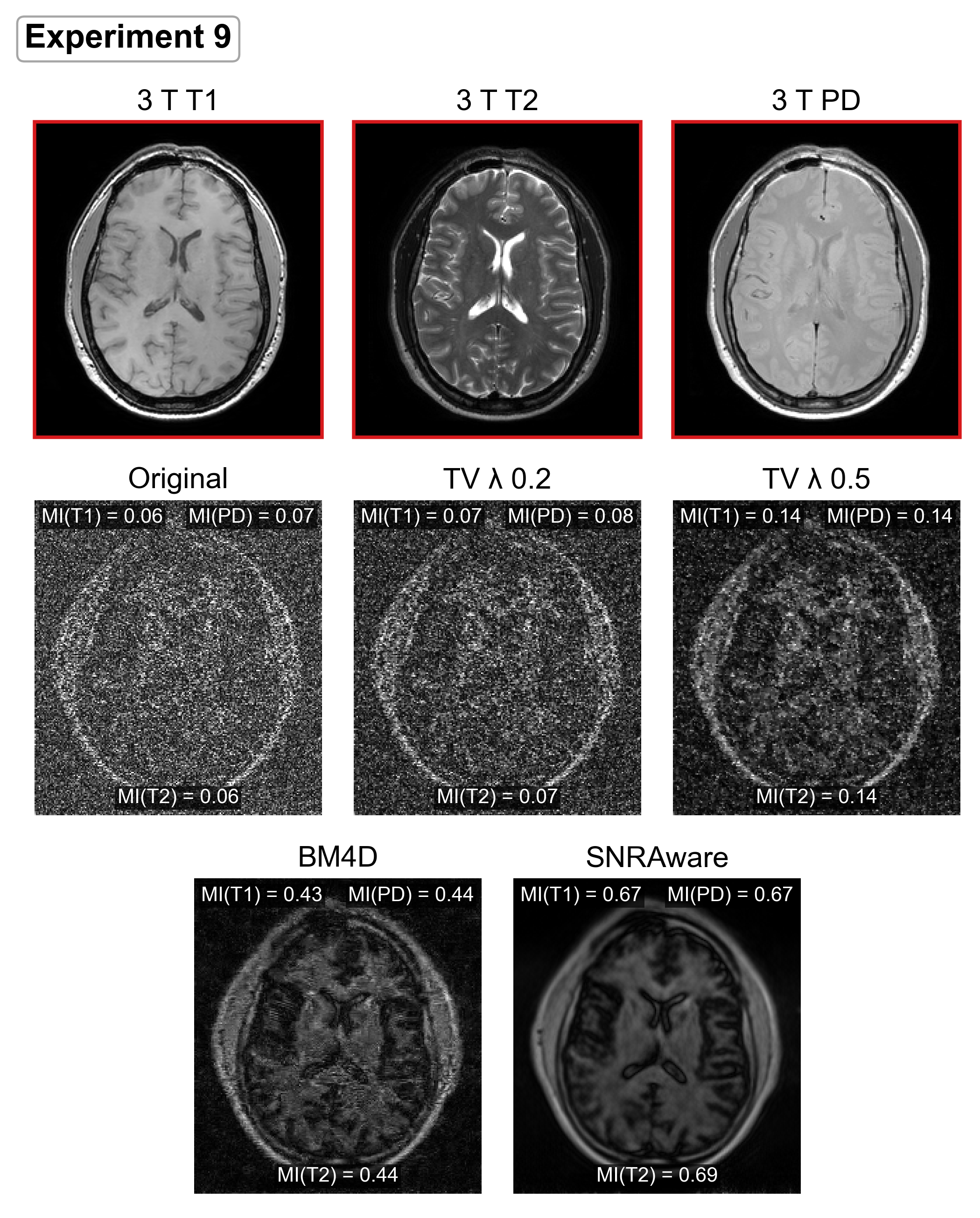}
	\caption{Experiment~9. ULF T$_1$-weighted brain images before and after denoising, compared with 3\,T T$_1$-weighted, T$_2$-weighted, and proton-density-weighted references (red frames). Mutual information is reported in bits.}
	\label{fig:Exp9}
\end{figure}

Figure~\ref{fig:Exp9} shows the 3\,T T$_1$-weighted, T$_2$-weighted, and proton-density-weighted references from left to right across the top row. The original ULF T$_1$-weighted image and the two TV outputs appear in the middle row, with BM4D and SNRAware below. Each ULF panel displays its MI with the three references, expressed in bits.

\section{Discussion}

\subsection{3\,T clinical-resolution benchmark}

The results of Experiment~1 show that SNRAware substantially increases the apparent ROI SNR of ULF images. In Fig.~\ref{fig:Exp1}, the apparent SNR increased from 4.75 to 55.48 for the knee at 88\,mT ($11.7\times$) and from 2.27 to 48.97 for the brain at 47\,mT ($21.6\times$). These were the largest enhancements among the methods compared. By reducing noise at these voxel sizes, denoising enables ULF imaging at nominal spatial resolutions commonly used in clinical 3\,T protocols. Importantly, comparable voxel sizes alone do not establish equivalent image quality, which also depends on tissue contrast, $B_0$ homogeneity, gradient linearity, and the SNR of the acquisition.

SNRAware yielded the lowest volume-averaged output-residual correlation in both acquisitions, with values of 0.02 for the knee and 0.05 for the brain. BM4D, however, yielded mean Moran's I values closer to zero. The two metrics capture different properties of the residual. The correlation measures linear association between the denoised image and the residual, while Moran's I measures spatial autocorrelation in the residual magnitude. A value close to zero for either metric alone is insufficient to establish that the removed information consists only of noise.

The central $k$-space slices shown in Fig.~\ref{fig:Exp1} feature strongly suppressed peripheral noise with SNRAware relative to the raw reconstruction, with pronounced high-frequency structure apparent after the denoising.

\subsection{Model performance versus spatial resolution}

Experiment~2 shows that SNRAware can improve the visibility of anatomical features across the range of voxel sizes tested in the sagittal knee. One possible explanation for the larger relative gains at finer resolutions is that anatomical features are represented at spatial scales closer to those encountered in the high-resolution clinical training images. The acquisitions also differed in starting SNR, and the 1\,mm acquisition used PF sampling, so the comparison does not isolate the effect of voxel size.

The quantitative SNR measurements reflect this trend. The apparent ROI SNR increased from 17.2 to 29.6 at 2\,mm ($1.7\times$), from 8.0 to 42.5 at approximately 1.5\,mm ($5.3\times$), and from 5.0 to 54.4 at 1\,mm ($10.9\times$). Thus, as voxel size decreased, the original SNR fell and the relative enhancement increased.

To explore the limits of this behavior, we attempted to push the acquisition to even higher spatial resolution using the ACR phantom. The experiment ultimately reached the hardware limits of the gradient amplifiers, which can deliver a maximum gradient strength of approximately 45\,mT/m, before the denoising model itself showed clear signs of failure. In this extreme case ($0.2\times0.2$\,mm$^2$ in-plane resolution), the apparent ROI SNR increased from 3.29 to 139.09 ($42.3\times$).

The phantom experiment provides a useful test of the qualitative behavior of SNRAware under ULF conditions. The model was trained on clinical 3\,T data and was not specifically trained to handle artifacts encountered at ULF, such as zipper artifacts produced by narrowband electromagnetic interference (EMI). Generative models can suppress or hallucinate features when they conflict with learned image priors \cite{Tivnan2024}. In the ACR phantom example in Fig.~\ref{fig:Exp2}, the smooth receive-chain correction in Eq.~\eqref{eq:noise_power_profile} leaves the EMI zipper visible in the input to SNRAware. The zipper remains after denoising, even as diffuse background fluctuations are strongly suppressed. Visual inspection of the complex residual magnitude showed a predominantly noise-like pattern, with no obvious imprint of the main phantom structures on the slice shown. The volume-averaged output-residual correlation and mean Moran's I were 0.04 and 0.00, respectively.

\subsection{Model performance versus readout bandwidth}
\label{sec:DisExp4}

The results of Experiment~3 are shown in Fig.~\ref{fig:Exps3-6}(a). Compared with the 12\,ms readouts, the high-bandwidth (2\,ms) readouts showed (i) lower apparent ROI SNR in the original images $\mathcal{I}$, (ii) more pronounced brightness modulation in the $-20$\,dB image despite nearly uniform background noise after correction, and (iii) a zipper artifact in the same image due to narrowband EMI pickup. A separate advantage of the shorter readout is (iv) its reduced sensitivity to off-resonance geometric distortion. SNRAware visibly reduced background fluctuations at both bandwidths, but the fine grids in the high-resolution region of the phantom were clearly resolved only with the longer readouts.

A plausible explanation for the remaining brightness modulation concerns the different frequency dependence of received signal and total noise. The stronger readout gradient in the 2\,ms acquisition maps the same phantom positions to approximately sixfold larger frequency offsets, extending farther into the tails of the coil's receive response. There, the received MR signal and coil thermal noise decrease, while receiver-added noise can remain appreciable. The correction in Eq.~\eqref{eq:profile_correction}, using the RF-off noise profile defined in Eq.~\eqref{eq:relative_noise_profile}, can therefore flatten background noise without fully compensating signal attenuation.

The fitted broadband floor power increased approximately 5.3-fold between readouts, close to the sixfold increase in receive bandwidth and consistent with an approximately constant noise spectral density. A resonant contribution superimposed on a constant amplifier voltage-noise floor has been observed experimentally in an NMR receive circuit \cite{Zavjalov2019}. Furthermore, amplifier-added noise can depend on the source impedance presented by the receive circuit \cite{Meys1978}. These observations support the possibility that received signal and total noise have different frequency dependences, such that the correction in Eq.~\eqref{eq:profile_correction} flattens background noise without fully compensating signal shading when the signal-bearing frequency range extends into the tails of the coil resonance.

\subsection{Model performance versus impedance matching}

The results of Experiment~4 are also shown in Fig.~\ref{fig:Exps3-6}(a). At both readout durations, optimized matching ($-20$\,dB) gave a higher apparent ROI SNR before denoising than degraded matching ($-3$\,dB). In the 12\,ms acquisitions, several fine grid patterns are more clearly resolved after denoising with optimized matching, although they are visible under both conditions. In the 2\,ms acquisitions, the fine grids remain poorly resolved under either matching condition.

The 2\,ms image acquired with degraded matching is more uniform in brightness. A plausible interpretation is that the broader effective receive response reduces the mismatch between signal attenuation and the total noise profile across the frequency range sampled by the phantom, as discussed in Sec.~\ref{sec:DisExp4}. The comparison therefore suggests a trade-off between the higher input SNR obtained with optimized matching and the more uniform brightness remaining after noise-profile correction with degraded matching.

\subsection{Model performance versus input noise level}

The effect of input noise amplitude in Experiment~5 is illustrated in Fig.~\ref{fig:Exps3-6}(b). SNRAware makes the main phantom structures easier to distinguish even when the original image is severely degraded by noise at $6.2\times$ the thermal level. However, the fine grids visible after denoising the $1.2\times$ thermal acquisition are largely indistinct in the elevated-noise output.

The apparent ROI SNR enhancement was $38.9\times$ in the baseline acquisition and $28.0\times$ in the elevated-noise acquisition, while the output-residual correlation increased from 0.02 to 0.04. However, the final apparent ROI SNR was 378.3 at baseline and 71.9 in the elevated-noise acquisition, and the zooms show that the visibility of weak structures after denoising still depends on the SNR of the input image.

\subsection{Model performance versus undersampling}

The effect of $k$-space undersampling in Experiment~6 is shown in Fig.~\ref{fig:Exps3-6}(c). SNRAware reduces background fluctuations in both the fully sampled and the 60\,\% PF acquisitions, and the main phantom structures remain visible in both outputs. In the high-resolution zooms, the grid on the left is more clearly resolved in the fully sampled image. In the PF image, some features remain visible as blurred patterns, with individual elements harder to distinguish.

The output-residual correlation was 0.02 for the fully sampled acquisition and 0.04 for the 60\,\% PF acquisition. The corresponding mean Moran's I values were 0.00 and 0.01, respectively. This comparison illustrates how visual inspection of the high-resolution region complements residual metrics averaged over the volume.

\subsection{Model performance versus anatomy, orientation, contrast, and scanner}

The results of Experiment~7 show that SNRAware provides substantial apparent SNR improvements across the anatomical targets, imaging orientations, contrasts, and scanners tested. In Fig.~\ref{fig:Exp7}, background fluctuations are visibly suppressed in the knee and wrist images acquired at 88\,mT, with anatomical structures becoming easier to distinguish in axial, sagittal, and coronal orientations. The carpal structures visible after denoising the wrist STIR acquisition with an inversion time of 80\,ms are particularly notable. The brain examples extend the comparison to the 47\,mT system, where cortical structures, sulcal CSF spaces, and the lateral ventricles are more readily distinguished after denoising in both T$_1$-weighted and T$_2$-weighted images. These observations support the applicability of the same SNRAware model across different anatomical structures and acquisition conditions, consistent with previous reports of ULF brain imaging \cite{Guallart-Naval2026b}.

The residual metrics nevertheless vary across the acquisitions. For example, the mean output-residual correlation is 0.04 for the axial T$_1$-weighted knee acquisition and 0.10 for the coronal STIR knee acquisition. The corresponding mean Moran's I values are 0.01 and 0.02, respectively. These differences illustrate that similar reductions in the visual prominence of noise can be accompanied by different degrees of linear association between the output and the removed information. Again, the range of performance across this experiment therefore reinforces the value of evaluating each acquisition using residual metrics alongside visual inspection.

\subsection{Model performance versus field strength}

The results of Experiment~8 compare ACR phantom acquisitions obtained with the NextMRI system at 88\,mT and the Physio~I scanner at 72\,mT. Both systems were operated after optimizing their hardware configuration to approach thermal-noise-limited conditions, providing a practical proxy for assessing denoising performance at different field strengths.

As shown in Fig.~\ref{fig:Exp8}, the high-resolution structures are more readily distinguished at 88\,mT, particularly in the original images. SNRAware substantially suppresses background fluctuations in both acquisitions, making these structures more apparent. The ROI SNR enhancements are $24.7\times$ at 88\,mT and $69.4\times$ at 72\,mT, the corresponding output-residual correlations are 0.03 and 0.04, and the mean Moran's I values are 0.00 and 0.01, respectively.

These observations are consistent with improved image quality at the higher field strength, although differences in RF coils and field homogeneity may also contribute. In particular, the greater geometric distortion visible in the NextMRI image is consistent with its higher field inhomogeneity; no distortion correction was applied in this comparison \cite{Borreguero2025,Guallart-Naval2026b}.

\subsection{Comparison with 3\,T clinical references}

Experiment~9 provides a complementary assessment of denoising by comparing the ULF images with independently acquired clinical references. SNRAware yielded the highest MI for all three reference contrasts, with values of 0.67-0.69\,bits, compared with 0.06-0.07\,bits for the original image, 0.07-0.08\,bits for TV with $\lambda=0.2$, 0.14\,bits for TV with $\lambda=0.5$, and 0.43-0.44\,bits for BM4D. This ranking was retained with 32 and 128 histogram bins, indicating that it was not specific to the selected binning. The greater statistical correspondence with the references accompanies the clearer delineation of ventricular and cortical structures in the SNRAware output.

The consistent ranking across T$_1$-weighted, proton-density-weighted, and T$_2$-weighted references supports the relevance of this comparison beyond a single contrast.

\section{Conclusions \& Outlook}

This work presented a systematic evaluation of DL-based denoising for ULF-MRI using a model trained with the SNRAware framework. Across a series of controlled experiments, the model demonstrated the ability to substantially increase the SNR of ULF acquisitions. In particular, Experiment~1 showed that knee images acquired at 88\,mT and brain images acquired at 47\,mT can reach nominal spatial resolutions comparable to those used in clinical 3\,T protocols, albeit without implying equivalence in overall image fidelity. The experiments further revealed that the denoising model performs consistently across a wide range of acquisition conditions, including different spatial resolutions, anatomical targets, contrasts, orientations, and scanner platforms. Comparison with independently acquired 3\,T brain references yielded the highest mutual information for SNRAware among the methods tested, providing complementary evidence of correspondence with the reference images without establishing equivalent anatomical fidelity.

A central observation emerging from this study is the ability of the model to remove noise while preserving the underlying signal structure. Residual images obtained by subtracting the denoised reconstructions from their corresponding inputs typically exhibit noise-like patterns with minimal structured content, indicating that the model acts predominantly as a noise suppressor rather than performing implicit image restoration or hallucination.

At the same time, the experiments highlight that denoising is not miraculous and is subject to several limitations. The quality of the denoised output remains dependent on the starting SNR of the acquisition, as illustrated by the elevated-noise experiment (Fig.~\ref{fig:Exps3-6}(b)). Likewise, the performance of the model appears to improve as the spatial resolution approaches the regime of the training data, suggesting that training-domain mismatch plays a role in determining the model’s behavior (Fig.~\ref{fig:Exp2}). Hardware factors such as impedance matching, bandwidth selection, and environmental noise also influence the final reconstruction quality, emphasizing that denoising performance ultimately reflects the combined properties of both the acquisition system and the post-processing algorithm. Proper sensitivity-profile correction and SNR-unit normalization are crucial, as departures from the expected input noise statistics can substantially alter denoising behavior.

The residual analysis provides a practical means of assessing this sensitivity. The relatively similar Moran's I values obtained for most SNRAware outputs are consistent with the care taken in sensitivity-profile correction and noise normalization, while the normalization experiments illustrate how this metric can flag deviations from the expected input noise statistics. Nevertheless, a low Moran's I does not by itself establish satisfactory denoising: visual inspection can reveal localized anatomy or artifacts that are poorly captured by a global measure. We conclude that quantitative residual metrics and visual assessment should therefore be used together.

Despite these limitations, the SNR gains observed in this work are sufficiently large to potentially alter the practical capabilities of ULF-MRI systems. Poor SNR has long been considered one of the primary limitations of ULF imaging, often competing with $B_0$ inhomogeneity as the dominant constraint on image quality \cite{Webb2023b,Ayde2025}. The results presented here indicate that DL-based denoising can mitigate this limitation to a considerable extent, enabling substantially higher spatial resolution imaging than previously achievable at these field strengths. It is therefore plausible that the principal bottleneck in ULF-MRI shifts from SNR to acquisition time, emphasizing the need for improved sampling efficiency and acceleration strategies.

Further progress will benefit from training models on datasets that explicitly include artifacts commonly observed in ULF systems, such as EMI patterns and receive-chain spectral modulation. Incorporating these effects into the training domain may allow DL-based denoisers to address both stochastic noise and structured acquisition artifacts, further expanding the applicability of these methods in ULF-MRI.

At the same time, the clinical implications of DL-based denoising must be rigorously assessed. The present study has taken a broad and systematic approach to characterizing the behavior of the model across a wide range of acquisition conditions, but the analysis remains focused on technical image properties. The ultimate test of these methods will therefore be their performance in clinical practice. This will require acquiring images from patient cohorts and evaluating them through blinded analysis by specialized radiologists in order to determine the true diagnostic value of DL-denoised ULF images. A first such study is presently under way in the scope of the NextMRI project \cite{NextMRI} and may establish whether the substantial SNR gains demonstrated here translate into meaningful improvements in clinical decision-making.

\appendices

\section*{Contributions}
See Table~\ref{tab:contributions}.

\begin{table*}
	\centering
	\fontsize{8.5}{9.5}\selectfont
	\caption{Author contributions. An “x” indicates participation in the corresponding task.}
	\label{tab:contributions}
	\begin{tabular}{l*{14}{c}}
		\toprule
		\textbf{Task} & \textbf{TGN} & \textbf{HX} & \textbf{JMA} &
		\textbf{EGC} & \textbf{JC} & \textbf{FG} & \textbf{CN} &
		\textbf{MAN} & \textbf{JS} & \textbf{RvdB} & \textbf{LVC} &
		\textbf{MH} & \textbf{AW} & \textbf{JA} \\
		\midrule
		Scanner preparation  &  x  &    &     &  x  & x  &    & x  &  x  &    &  x   &  x  &    & x  & x  \\
		SNRAware/Tyger prep. &     & x  &     &     &    &    &    &     & x  &      &     & x  &    &    \\
		Data acq.            &  x  &    &  x  &  x  & x  &    &    &  x  &    &      &  x  &    &    & x  \\
		Data processing      &  x  & x  &     &     &    &    &    &     &    &      &     &    &    & x  \\
		Proj. conception     &  x  & x  &     &     &    &    &    &     &    &      &     & x  & x  & x  \\
		Proj. manage.        &     &    &     &     &    & x  &    &     &    &      &     & x  & x  & x  \\
		Figure prod.         &  x  & x  &  x  &     &    &    &    &     &    &      &     &    &    & x  \\
		Paper writing        &  x  & x  &     &     &    &    &    &     &    &      &     &    & x  & x  \\
		Paper revision       &  x  & x  &  x  &  x  & x  & x  & x  &  x  & x  &  x   &  x  & x  & x  & x  \\
		\bottomrule
	\end{tabular}
	\normalsize
\end{table*}

\section*{Funding}
This work was supported by Ministerio de Ciencia e Innovación (PID2022-142719OB-C22), the European Innovation Council (NextMRI 101136407), and the Wellcome Trust (317797/Z/24/Z). JC acknowledges funding from the Spanish Ministry of Science, Innovation and Universities through a Formación del Profesorado Universitario grant (FPU23/01559).

\section*{Ethical statement}
The \emph{in vivo} experiments performed at i3M were approved by the Ethics Committee of the Spanish National Research Council (research agreement number 276/2025). The study adhered to the guidelines of the Leiden University Medical Center Institutional Review Board (The Netherlands). Informed consent to participate and for publication was obtained from the volunteers prior to study commencement.

\section*{Code and data availability}
The SNRAware framework and its publicly released models are available at \url{https://github.com/microsoft/SNRAware/}. SNR-ND, the open-source tool used for reconstruction and normalization in this work, is available at \url{https://github.com/tj-rnd/SNR-ND} \cite{SNRND2026}. The MRD acquisitions, reference images, denoised outputs, and processing metadata supporting this study are available at \url{https://huggingface.co/datasets/tj-rnd/snraware-ulf-mri_paper}.

\section*{Conflict of interest}
JMA, FG, and JA are co-founders of PhysioMRI Tech.

\ifCLASSOPTIONcaptionsoff
  \newpage
\fi


\begin{thebibliography}{10}
\providecommand{\url}[1]{#1}
\csname url@samestyle\endcsname
\providecommand{\newblock}{\relax}
\providecommand{\bibinfo}[2]{#2}
\providecommand{\BIBentrySTDinterwordspacing}{\spaceskip=0pt\relax}
\providecommand{\BIBentryALTinterwordstretchfactor}{4}
\providecommand{\BIBentryALTinterwordspacing}{\spaceskip=\fontdimen2\font plus
\BIBentryALTinterwordstretchfactor\fontdimen3\font minus
  \fontdimen4\font\relax}
\providecommand{\BIBforeignlanguage}[2]{{%
\expandafter\ifx\csname l@#1\endcsname\relax
\typeout{** WARNING: IEEEtran.bst: No hyphenation pattern has been}%
\typeout{** loaded for the language `#1'. Using the pattern for}%
\typeout{** the default language instead.}%
\else
\language=\csname l@#1\endcsname
\fi
#2}}
\providecommand{\BIBdecl}{\relax}
\BIBdecl

\bibitem{AjaBookNoise}
S.~Aja-Fern{\'a}ndez and G.~Vegas-S{\'a}nchez-Ferrero, \emph{Statistical
  Analysis of Noise in {MRI}}.\hskip 1em plus 0.5em minus 0.4em\relax Springer,
  2016.

\bibitem{Webb2023b}
A.~Webb and T.~O’Reilly, ``{Tackling SNR at low-field: a review of hardware
  approaches for point-of-care systems},'' \emph{Magnetic Resonance Materials
  in Physics, Biology and Medicine}, vol.~36, no.~3, pp. 375--393, 2023.

\bibitem{Biber2025}
\BIBentryALTinterwordspacing
S.~Biber, S.~Kannengiesser, J.~Nistler, M.~Braun, S.~Blaess, M.~Gebhardt,
  D.~Grodzki, D.~Ritter, G.~Seegerer, M.~Vester, and R.~Schneider, ``{Design
  and operation of a whole-body MRI scanner without RF shielding},''
  \emph{{Magnetic Resonance in Medicine}}, vol.~93, no.~4, pp. 1842--1855,
  2025. [Online]. Available:
  \url{https://onlinelibrary.wiley.com/doi/abs/10.1002/mrm.30374}
\BIBentrySTDinterwordspacing

\bibitem{Bellon86}
\BIBentryALTinterwordspacing
E.~Bellon, E.~Haacke, P.~Coleman, D.~Sacco, D.~Steiger, and R.~Gangarosa, ``{MR
  artifacts: a review},'' \emph{{American Journal of Roentgenology}}, vol. 147,
  no.~6, pp. 1271--1281, 1986, pMID: 3490763. [Online]. Available:
  \url{https://doi.org/10.2214/ajr.147.6.1271}
\BIBentrySTDinterwordspacing

\bibitem{Guallart-Naval2026a}
\BIBentryALTinterwordspacing
T.~Guallart-Naval, J.~M. Algarín, and J.~Alonso, ``{Electromagnetic Noise
  Characterization and Suppression in Low-Field MRI Systems},'' \emph{{Magnetic
  Resonance in Medicine}}, vol.~95, no.~5, pp. 3000--3007, 2026. [Online].
  Available: \url{https://onlinelibrary.wiley.com/doi/abs/10.1002/mrm.70235}
\BIBentrySTDinterwordspacing

\bibitem{BkHaacke}
E.~M. Haacke, R.~W. Brown, M.~R. Thompson, R.~Venkatesan \emph{et~al.},
  \emph{Magnetic resonance imaging: physical principles and sequence
  design}.\hskip 1em plus 0.5em minus 0.4em\relax Wiley-liss New York:, 1999,
  vol.~82.

\bibitem{Manjon2012}
\BIBentryALTinterwordspacing
J.~V. Manjón, P.~Coupé, A.~Buades, D.~{Louis Collins}, and M.~Robles, ``{New
  methods for MRI denoising based on sparseness and self-similarity},''
  \emph{Medical Image Analysis}, vol.~16, no.~1, pp. 18--27, 2012. [Online].
  Available:
  \url{https://www.sciencedirect.com/science/article/pii/S1361841511000491}
\BIBentrySTDinterwordspacing

\bibitem{Mohan2014}
\BIBentryALTinterwordspacing
J.~Mohan, V.~Krishnaveni, and Y.~Guo, ``A survey on the magnetic resonance
  image denoising methods,'' \emph{Biomedical Signal Processing and Control},
  vol.~9, pp. 56--69, 2014. [Online]. Available:
  \url{https://www.sciencedirect.com/science/article/pii/S1746809413001407}
\BIBentrySTDinterwordspacing

\bibitem{Kyung2018}
A.~Kyung, Y.~J. Kwon, and L.~Chung, ``Effects of k-space spatial low-pass
  filtering on bio-imaging analysis,'' in \emph{2018 9th IEEE Annual Ubiquitous
  Computing, Electronics \& Mobile Communication Conference (UEMCON)}, 2018,
  pp. 742--745.

\bibitem{Wang2025}
\BIBentryALTinterwordspacing
J.~Wang and J.~P. Haldar, ``{Well-Designed k-Space Coverage is Important for
  Good MRI Denoising},'' \emph{arXiv preprint arXiv:2511.05735}, 2025.
  [Online]. Available: \url{https://arxiv.org/abs/2511.05735}
\BIBentrySTDinterwordspacing

\bibitem{Ciulla2025}
\BIBentryALTinterwordspacing
C.~Ciulla, ``{Two-dimensional image noise removal and reconstruction using
  discrete Fourier transform, k-space filtering and Z-space filtering},''
  \emph{Progress in Engineering Science}, vol.~2, no.~1, p. 100056, 2025.
  [Online]. Available:
  \url{https://www.sciencedirect.com/science/article/pii/S2950425225000088}
\BIBentrySTDinterwordspacing

\bibitem{Manjon2018}
J.~V. Manj{\'o}n and P.~Coupe, ``{MRI denoising using deep learning},'' in
  \emph{International workshop on patch-based techniques in medical
  imaging}.\hskip 1em plus 0.5em minus 0.4em\relax Springer, 2018, pp. 12--19.

\bibitem{Ongie2020}
G.~Ongie, A.~Jalal, C.~A. Metzler, R.~G. Baraniuk, A.~G. Dimakis, and
  R.~Willett, ``Deep learning techniques for inverse problems in imaging,''
  \emph{IEEE Journal on Selected Areas in Information Theory}, vol.~1, no.~1,
  pp. 39--56, 2020.

\bibitem{Xue2025}
\BIBentryALTinterwordspacing
H.~Xue, S.~M. Hooper, I.~Pierce, R.~H. Davies, J.~Stairs, J.~Naegele, A.~E.
  Campbell-Washburn, C.~Manisty, J.~C. Moon, T.~A. Treibel, M.~S. Hansen, and
  P.~Kellman, ``{SNRAware: Improved Deep Learning MRI Denoising with
  Signal-to-Noise Ratio Unit Training and G-Factor Map Augmentation},''
  \emph{Radiology: Artificial Intelligence}, vol.~7, no.~6, p. e250227, 2025,
  pMID: 41123451. [Online]. Available:
  \url{https://doi.org/10.1148/ryai.250227}
\BIBentrySTDinterwordspacing

\bibitem{Xue2025b}
\BIBentryALTinterwordspacing
H.~Xue, S.~M. Hooper, R.~H. Davies, T.~A. Treibel, I.~Pierce, J.~Stairs,
  J.~Naegele, C.~Manisty, J.~C. Moon, A.~E. Campbell-Washburn, P.~Kellman, and
  M.~S. Hansen, ``{Imaging Transformer for MRI Denoising: a Scalable Model
  Architecture that enables SNR $\ll$ 1 Imaging},'' \emph{{arXiv preprint
  arXiv: 2504.10534}}, 2025. [Online]. Available:
  \url{https://arxiv.org/abs/2504.10534}
\BIBentrySTDinterwordspacing

\bibitem{Webb2023}
\BIBentryALTinterwordspacing
A.~Webb and J.~Obungoloch, ``{Five steps to make MRI scanners more affordable
  to the world},'' \emph{{Nature}}, vol. 615, pp. 391--393, 3 2023. [Online].
  Available: \url{https://www.nature.com/articles/d41586-023-00759-x}
\BIBentrySTDinterwordspacing

\bibitem{Guallart-Naval2022}
\BIBentryALTinterwordspacing
T.~Guallart-Naval, J.~M. Algar{\'{i}}n, R.~Pellicer-Guridi, F.~Galve,
  Y.~Vives-Gilabert, R.~Bosch, E.~Pall{\'{a}}s, J.~M. Gonz{\'{a}}lez, J.~P.
  Rigla, P.~Mart{\'{i}}nez, F.~Lloris, J.~Borreguero, {\'{A}}.~Marcos-Perucho,
  V.~Negnevitsky, L.~Mart{\'{i}}-Bonmat{\'{i}}, A.~R{\'{i}}os, J.~M. Benlloch,
  and J.~Alonso, ``{Portable magnetic resonance imaging of patients indoors,
  outdoors and at home},'' \emph{Scientific Reports 2022 12:1}, vol.~12, no.~1,
  pp. 1--11, jul 2022. [Online]. Available:
  \url{https://www.nature.com/articles/s41598-022-17472-w}
\BIBentrySTDinterwordspacing

\bibitem{Zhao2024}
\BIBentryALTinterwordspacing
Y.~Zhao, Y.~Ding, V.~Lau, C.~Man, S.~Su, L.~Xiao, A.~T.~L. Leong, and E.~X. Wu,
  ``{Whole-body magnetic resonance imaging at 0.05 tesla},'' \emph{Science},
  vol. 384, 5 2024. [Online]. Available:
  \url{https://www.science.org/doi/10.1126/science.adm7168}
\BIBentrySTDinterwordspacing

\bibitem{Sarracanie2020}
\BIBentryALTinterwordspacing
M.~Sarracanie and N.~Salameh, ``{Low-Field MRI: How Low Can We Go? A Fresh View
  on an Old Debate},'' \emph{Frontiers in Physics}, vol.~8, p. 172, jun 2020.
  [Online]. Available:
  \url{https://www.frontiersin.org/article/10.3389/fphy.2020.00172/full}
\BIBentrySTDinterwordspacing

\bibitem{Ayde2025}
R.~Ayde, M.~Vornehm, Y.~Zhao, F.~Knoll, E.~X. Wu, and M.~Sarracanie, ``{MRI at
  low field: A review of software solutions for improving SNR},'' \emph{NMR in
  Biomedicine}, vol.~38, no.~1, p. e5268, 2025.

\bibitem{Maggioni2013}
M.~Maggioni, V.~Katkovnik, K.~Egiazarian, and A.~Foi, ``{Nonlocal
  transform-domain filter for volumetric data denoising and reconstruction},''
  \emph{IEEE Transactions on Image Processing}, vol.~22, no.~1, pp. 119--133,
  2013.

\bibitem{Liu2021}
\BIBentryALTinterwordspacing
Y.~Liu, A.~T.~L. Leong, Y.~Zhao, L.~Xiao, H.~K.~F. Mak, A.~Chun, O.~Tsang,
  G.~K.~K. Lau, G.~K.~K. Leung, E.~X. Wu, and X.~Linfang, ``{A low-cost and
  shielding-free ultra-low-field brain MRI scanner},'' \emph{Nature
  Communications 2021 12:1}, vol.~12, no.~1, pp. 1--14, dec 2021. [Online].
  Available: \url{https://www.nature.com/articles/s41467-021-27317-1}
\BIBentrySTDinterwordspacing

\bibitem{Zhao2024b}
\BIBentryALTinterwordspacing
Y.~Zhao, L.~Xiao, J.~Hu, and E.~X. Wu, ``Robust emi elimination for rf
  shielding-free mri through deep learning direct mr signal prediction,''
  \emph{Magnetic Resonance in Medicine}, vol.~92, no.~1, pp. 112--127, 2024.
  [Online]. Available:
  \url{https://onlinelibrary.wiley.com/doi/abs/10.1002/mrm.30046}
\BIBentrySTDinterwordspacing

\bibitem{Salehi2025}
\BIBentryALTinterwordspacing
A.~Salehi, M.~Mach, C.~Najac, B.~Lena, T.~O’Reilly, Y.~Dong, P.~Börnert,
  H.~Adams, T.~Evans, and A.~Webb, ``{Denoising low-field MR images with a deep
  learning algorithm based on simulated data from easily accessible open-source
  software},'' \emph{Journal of Magnetic Resonance}, vol. 370, p. 107812, 2025.
  [Online]. Available:
  \url{https://www.sciencedirect.com/science/article/pii/S1090780724001964}
\BIBentrySTDinterwordspacing

\bibitem{Ilicak2025}
\BIBentryALTinterwordspacing
E.~Ilıcak, C.~Rao, C.~Najac, B.~Lena, B.~Imre, F.~Galve, J.~Alonso, A.~Webb,
  and M.~Staring, ``{Physics-Informed Deep Unrolled Network for Portable MR
  Image Reconstruction},'' \emph{arXiv preprint arXiv:2509.11790}, 2025.
  [Online]. Available: \url{https://arxiv.org/abs/2509.11790}
\BIBentrySTDinterwordspacing

\bibitem{Kellman2005}
P.~Kellman and E.~R. McVeigh, ``{Image reconstruction in SNR units: A general
  method for SNR measurement},'' \emph{Magnetic Resonance in Medicine},
  vol.~54, no.~6, pp. 1439--1447, 2005.

\bibitem{Meys1978}
R.~P. Meys, ``A wave approach to the noise properties of linear microwave
  devices,'' \emph{IEEE Transactions on Microwave Theory and Techniques},
  vol.~26, no.~1, pp. 34--37, 1978.

\bibitem{Zavjalov2019}
V.~V. Zavjalov, A.~M. Savin, and P.~J. Hakonen, ``Cryogenic differential
  amplifier for {NMR} applications,'' \emph{Journal of Low Temperature
  Physics}, vol. 195, pp. 72--80, 2019.

\bibitem{Brunet2009}
D.~Brunet, E.~R. Vrscay, and Z.~Wang, ``The use of residuals in image
  denoising,'' in \emph{Image Analysis and Recognition}, ser. Lecture Notes in
  Computer Science, M.~Kamel and A.~Campilho, Eds.\hskip 1em plus 0.5em minus
  0.4em\relax Berlin, Heidelberg: Springer, 2009, vol. 5627, pp. 1--12.

\bibitem{Moran1950}
P.~A.~P. Moran, ``Notes on continuous stochastic phenomena,''
  \emph{Biometrika}, vol.~37, no. 1--2, pp. 17--23, 1950.

\bibitem{Galve2024}
\BIBentryALTinterwordspacing
F.~Galve, E.~Pallás, T.~Guallart-Naval, P.~García-Cristóbal, P.~Martínez,
  J.~M. Algarín, J.~Borreguero, R.~Bosch, F.~Juan-Lloris, J.~M. Benlloch, and
  J.~Alonso, ``Elliptical halbach magnet and gradient modules for low-field
  portable magnetic resonance imaging,'' \emph{NMR in Biomedicine}, vol.~37,
  no.~12, p. e5258, 2024. [Online]. Available:
  \url{https://analyticalsciencejournals.onlinelibrary.wiley.com/doi/abs/10.1002/nbm.5258}
\BIBentrySTDinterwordspacing

\bibitem{Lena2026}
B.~Lena, R.~van~den Broek, T.~O'Reilly, A.~Webb, and C.~Najac, ``Evaluating
  repeatability and reproducibility of in-vivo imaging using a portable
  {Halbach}-based 47\,mt scanner,'' 2026.

\bibitem{Negnevitsky2023}
\BIBentryALTinterwordspacing
V.~Negnevitsky, Y.~Vives-Gilabert, J.~M. Algar\'in, L.~Craven-Brightman,
  R.~Pellicer-Guridi, T.~O'Reilly, J.~P. Stockmann, A.~Webb, J.~Alonso, and
  B.~Menk\"uc, ``{MaRCoS, an open-source electronic control system for
  low-field MRI},'' \emph{Journal of Magnetic Resonance}, vol. 350, p. 107424,
  2023. [Online]. Available: \url{https://doi.org/10.1016/j.jmr.2023.107424}
\BIBentrySTDinterwordspacing

\bibitem{Guallart-Naval2022b}
T.~Guallart-Naval, T.~O'Reilly, J.~M. Algar\'in, R.~Pellicer-Guridi,
  Y.~Vives-Gilabert, L.~Craven-Brightman, V.~Negnevitsky, B.~Menk\"uc,
  F.~Galve, J.~P. Stockmann, A.~Webb, and J.~Alonso, ``{Benchmarking the
  performance of a low‐cost magnetic resonance control system at multiple
  sites in the open MaRCoS community},'' \emph{{NMR in Biomedicine}}, vol.~36,
  pp. 1--13, 2022.

\bibitem{Algarin2024}
J.~M. Algar{\'\i}n, T.~Guallart-Naval, J.~Borreguero, F.~Galve, and J.~Alonso,
  ``{MaRGE: A graphical environment for MaRCoS},'' \emph{{Journal of Magnetic
  Resonance}}, vol. 361, p. 107662, 2024.

\bibitem{SNRND2026}
\BIBentryALTinterwordspacing
J.~Alonso and T.~Guallart-Naval, ``{SNR-ND}: Snr-unit normalization and
  denoising for cartesian {MRI},'' GitHub software repository, 2026, version
  0.1.0.dev4, commit 04c9491. [Online]. Available:
  \url{https://github.com/tj-rnd/SNR-ND}
\BIBentrySTDinterwordspacing

\bibitem{Tyger}
\BIBentryALTinterwordspacing
``Github - microsoft/tyger: Remote signal processing.'' [Online]. Available:
  \url{https://github.com/microsoft/tyger}
\BIBentrySTDinterwordspacing

\bibitem{Guallart-Naval2026b}
T.~Guallart-Naval, J.~Stairs, J.~M. Algarín, H.~Xue, J.~Benlloch, P.~Benlloch,
  J.~Borreguero, J.~Conejero, F.~Galve, P.~García-Cristóbal, M.~Lacalle,
  B.~Lena, L.~Porcar, S.~J. Schiff, A.~Webb, M.~Hansen, and J.~Alonso, ``{A
  fully open-source framework for streaming and cloud-processing of low-field
  MRI data},'' \emph{{NMR in Biomedicine}}, 2026.

\bibitem{Haacke1991}
E.~M. Haacke, E.~D. Lindskog, and W.~Lin, ``A fast, iterative, partial-fourier
  technique capable of local phase recovery,'' \emph{Journal of Magnetic
  Resonance}, vol.~92, no.~1, pp. 126--145, 1991.

\bibitem{Uecker2015}
\BIBentryALTinterwordspacing
M.~Uecker, F.~Ong, J.~I. Tamir, D.~Bahri, P.~Virtue, J.~Y. Cheng, T.~Zhang, and
  M.~Lustig, ``Berkeley advanced reconstruction toolbox,'' in \emph{Proceedings
  of the 23rd Annual Meeting of the International Society for Magnetic
  Resonance in Medicine}, 2015, p. 2486. [Online]. Available:
  \url{https://archive.ismrm.org/2015/2486.html}
\BIBentrySTDinterwordspacing

\bibitem{Ghazinoor2007}
\BIBentryALTinterwordspacing
S.~Ghazinoor, J.~V. Crues, and C.~Crowley, ``{Low-field musculoskeletal MRI},''
  \emph{{Journal of Magnetic Resonance Imaging}}, vol.~25, pp. 234--244, 2
  2007. [Online]. Available: \url{http://doi.wiley.com/10.1002/jmri.20854}
\BIBentrySTDinterwordspacing

\bibitem{Maes1997}
F.~Maes, A.~Collignon, D.~Vandermeulen, G.~Marchal, and P.~Suetens,
  ``Multimodality image registration by maximization of mutual information,''
  \emph{IEEE Transactions on Medical Imaging}, vol.~16, no.~2, pp. 187--198,
  1997.

\bibitem{Shannon1948}
C.~E. Shannon, ``A mathematical theory of communication,'' \emph{The Bell
  System Technical Journal}, vol.~27, no.~3, pp. 379--423, 1948.

\bibitem{Tivnan2024}
M.~Tivnan, S.~Yoon, Z.~Chen, X.~Li, D.~Wu, and Q.~Li, ``{Hallucination Index:
  An Image Quality Metric for Generative Reconstruction Models},'' in
  \emph{{Medical Image Computing and Computer Assisted Intervention -- MICCAI
  2024}}, M.~G. Linguraru, Q.~Dou, A.~Feragen, S.~Giannarou, B.~Glocker,
  K.~Lekadir, and J.~A. Schnabel, Eds.\hskip 1em plus 0.5em minus 0.4em\relax
  Springer Nature Switzerland, 2024, pp. 449--458.

\bibitem{Borreguero2025}
\BIBentryALTinterwordspacing
J.~Borreguero, F.~Galve, J.~M. Algarín, and J.~Alonso, ``Zero-echo-time
  sequences in highly inhomogeneous fields,'' \emph{Magnetic Resonance in
  Medicine}, vol.~93, no.~3, pp. 1190--1204, 2025. [Online]. Available:
  \url{https://onlinelibrary.wiley.com/doi/abs/10.1002/mrm.30352}
\BIBentrySTDinterwordspacing

\bibitem{NextMRI}
\BIBentryALTinterwordspacing
{European Commission}, ``{NextMRI}: Truly portable {MRI} for extremity and
  brain imaging anywhere \& everywhere,'' CORDIS project fact sheet, 2023,
  grant agreement No. 101136407. Accessed: 2026-11-09. [Online]. Available:
  \url{https://cordis.europa.eu/project/id/101136407}
\BIBentrySTDinterwordspacing

\end{thebibliography}

\end{document}